\documentclass[fleqn,usenatbib]{mnras}

\usepackage{newtxtext,newtxmath}
\usepackage{adjustbox}
\usepackage{dblfloatfix}
\usepackage{lscape}
\usepackage{booktabs}
\usepackage[T2a,T1]{fontenc}
\usepackage{auto-pst-pdf}
\usepackage{graphicx}	
\usepackage{etoolbox,xpatch}
\usepackage{amsmath}	
\usepackage{xcolor}
\usepackage{color}
\usepackage{ulem}
\usepackage{comment}
\usepackage{pdflscape}
\usepackage{longtable}
\usepackage[english]{babel}
\usepackage{adjustbox}
\usepackage[skip=0pt]{caption}
\usepackage{graphicx}
\usepackage{epstopdf}
\newcommand\rem[1]{}

\title{The 2MIG isolated AGNs – 2. X-ray general properties and peculiarities}

\author[O. Kompaniiets et al.]{
O. V. Kompaniiets,$^{1}$\thanks{E-mail: kompaniets@mao.kiev.ua}
A. A. Vasylenko,$^{1}$
I. B. Vavilova$^{1}$
\\
$^{1}$Department for Extragalactic Astronomy and Astroinformatics, Main Astronomical Observatory of the National Academy of Sciences of Ukraine \\27, Akademik Zabolotnyi St., Kyiv 03123, Ukraine
}

\date{Accepted 2026 May 28. Received 2026 May 28; in original form 2025 May 02.}

\pubyear{2026}

\begin{document}
\label{firstpage}
\pagerange{\pageref{firstpage}--\pageref{lastpage}}
\maketitle

\begin{abstract}
We have analysed a sample of 2MIG isolated galaxies hosting AGNs (isolated AGNs) to assess whether their nuclear activity differs from that in denser environments. The isolation criteria rule out interactions with other galaxies of similar evolutionary stage at least $\sim 3$~Gyr. We systematised the available \textit{Swift}, \textit{NuSTAR}, \textit{XMM-Newton}, \textit{Chandra}, and \textit{INTEGRAL} X-ray data for isolated AGNs at $z<0.05$, determining their general X-ray properties and peculiarities, spectral models, and supermassive black hole (SMBH) masses. We investigated the best spectral models for 20 isolated AGNs, including 9 sources fitted for the first time in this work. Our results indicate that an isolation of galaxies in the nearby Universe does not significantly affect nuclear activity. This conclusion is supported by the diversity of accretion types and the absence of any preference for a particular basic or composite spectral model. We note an interesting case for ESO~499$-$041, where the data-to-model ratio shows significant changes above 5~keV. These deviations from the continuum may tentatively indicate the presence of a relativistic iron line in the \textit{Chandra} spectrum of ESO~499$-$041 compared to the \textit{Swift} spectrum. We derived SMBH masses for 24/32 isolated AGNs applying the $M_{\rm BH}$--$\sigma_\star$ relation and used available estimates for 8/32. A general distribution is that 27/32 isolated AGNs host SMBHs with $M_{\mathrm{SMBH}} \lesssim 10^{8}\,M_\odot$. For the first time, we found evidence of the linear correlation between L$_{2-10, keV}$ and $\log M_{\mathrm{SMBH}}$ that is not observed in other AGNs samples. This trend may be affected by the limited sample size and requires further confirmation with larger datasets. 
\end{abstract} 


\section{Introduction}

Isolated galaxies are unique laboratories for studying their internal properties in the low-density intergalactic medium. Their location in or near cosmic voids significantly reduces the probability of recent interactions, preserving their evolutionary history over the past few billion years. This makes isolation a principal criterion for their classification \citep{Marguez1999, 2007Verley, Nigoshe2007,  Vavilova2009, Wang2012}. The first Catalog of Isolated Galaxies (CIG) was developed by \cite{1973Karachentseva}. Since then, the number of identified isolated systems has increased, along with the variety of criteria used to determine the isolation (e.g., \citep{Elyiv2009, Karachentseva2010, Fernandez2013, Spector2017, Spector2020, Vavilova2021, Argudo2025, Benavides2025}). These improvements have enabled a more detailed analysis of their physical properties and the relationships between them, such as the stellar mass-to-size ratio \citep{2013MNRASFernandezLorenzo}, the scaling of the mass H\,\textsc{i} with the optical diameter D$_{25}$ \citep{2018AAJones}, the HI scaling relation for isolated galaxies using WISE stellar masses \citep{Bok2020MNRAS} or the lowest level of asymmetry in their integrated HI profiles \citep{Moreta2018}, and other properties established, for example, in the framework of the AMIGA \citep{Sorgho2024} and CAVITY \citep{Perez2024} projects.

Cosmological simulations \citep{Hirschmann2013MNRAS, Spector2016MNRAS, Habouzit2020MNRAS} show that galaxy evolution is influenced by multiple factors, including internal properties and the environments in which galaxies reside. 

\cite{Kinyumu2024MNRAS} studied properties of isolated galaxies in the Digital Survey Isolated Galaxies (DSIG) catalogue within a redshift range of 0.005 < z < 0.08. The concentration index, Cr = 2.65, divides their sample of isolated galaxies into early and late types, with late types (spiral galaxies) comprising 68 \% of the isolated sample. \cite{Miraghaei2020} examined active galactic nuclei (AGN) in different environments. The results obtained in this work confirm the absence of environmental dependence of optical AGN activity in blue galaxies and its lower significance in green galaxies, whereas in massive red galaxies in voids, a higher fraction of optical AGNs is observed than in galaxies in dense environments. The radio activity of these AGNs increases in the dense environments of red galaxies. No changes in the radio-loud AGN fraction are observed for blue and green galaxies. This indicates that environmental effects on AGN activity are not significant in galaxies with cold gas.

\cite{Pulatova2015} analysed multiwavelength properties of isolated 2MIG galaxies with AGNs at $z$<0.1 from the 2MIG catalogue \citep{Karachentseva2010} and found that host galaxies are of morphological types S0-Sc, and half of these AGNs are classified as Seyfert 2 (Sy2). They also revealed that the host galaxies of AGNs of the Sy1 type (without faint companions) appear to possess bar-like morphological features. This suggests that a bar is crucial to the existence of a broad-line region (BLR) in an isolated host galaxy, providing a transfer of gas and dust from the galaxy's disc to the AGN region. This leads to the conclusion that interaction with neighbouring galaxies is not necessary for BLR formation.

Isolated AGNs are predominantly faint across the radio-to-X-ray bands. In the radio band, \cite{2023KFNTPulatova} found that 51 2MIG isolated AGNs exhibit radio-quiet characteristics. Most of them have flux densities at a frequency of 1.4 GHz within the range of 3-20 mJy, as well as less than 3 mJy for some isolated AGNs. Exceptions are related to PGC 35009 and NGC 6951 with flux densities in the range of 50-200 mJy, and ESO 483-009 and ESO 097-013 with 352 and 1200 mJy, respectively. X-ray emission is a key indicator of galactic nucleus activity. Spectral models and parameter values (spectral index and intrinsic absorption) were determined for several isolated AGNs: CGCG 179-005, NGC 6300, NGC 1050, WKK 3050, ESO 438-009, and ESO 317-038 \citep{Chesnok2010, Vavilova2015, Pulatova2015}. X-ray spectra for bright galaxies, notably NGC 6300 and Circinus, were examined up to 250 keV, with particular emphasis on emission features in the 6-7 keV range. Additional insights by \cite{Vavilova2016, Vasylenko2020} indicated that isolated AGNs typically possess smaller luminosities (L$_{2-10 keV}$ $\sim$ $10^{42}$ erg/s) compared to typical Seyfert galaxies. NGC 5347 and MCG-02-09-040 spectra exhibited neutral Fe~K${\alpha}$ emission lines, with NGC 5347 described by a pure reflection model, and MCG-02-09-040 showed a heavy neutral obscuration. 

In our previous study of the multiwavelength properties of 2MIG isolated AGNs \citep{kompaniiets2023knit}, we found that the contribution of AGN to the total emission of the galaxy at this evolutionary stage does not exceed 10$\%$, and more than two-thirds of active nuclei do not emit in the X-ray range. \cite{2024KosNTVavilova} emphasised that the isolation and faint nuclear activity criteria should be among the key parameters in the search for Milky Way galaxy analogues (MWAs). In this context, isolated AGNs represent an important evolutionary class that can help us better understand the formation and evolutionary pathway of our Galaxy. \citet{Sivasankaran2025} investigated the role of AGN feedback in isolated galaxies using high-resolution AREPO simulations with the SMUGGLE model, which accounts for multiphase interstellar medium and stochastic star formation. Their models included three representative galaxy types: a thin-disc MWAs, a gas-rich thick-disc Small Magellanic Cloud (SMC)-type dwarf galaxy, and a luminous infrared Sbc-type galaxy. Their study demonstrated that AGN-driven winds effectively regulate supermassive black hole (SMBH) growth by heating and expelling gas from the nuclear regions, with the efficiency of feedback depending strongly on galaxy morphology. In thin-disc systems such as the MWAs, AGN energy predominantly escapes vertically with limited interaction with the disc, resulting in minimal impact on star formation. In contrast, in galaxies with thicker gas discs (Sbc and SMC types), AGN feedback couples more efficiently with the interstellar medium (ISM), driving powerful outflows and significantly suppressing star formation up to 70$\%$ in the SMC-type galaxy.

Our research aims to determine the general X-ray properties and peculiarities of 2MIG isolated AGNs at z < 0.05 using available observational data, to explore spectral models for various AGN types, to estimate SMBH masses, and to define relationships among the main X-ray parameters of isolated AGNs. 

The paper is structured as follows. In Section \ref{sec:sample}, we describe the galaxy sample. Section \ref{sec:Xray_generall} presents the X-ray analysis and summarises the general X-ray properties. In Section \ref{sec:SMBHmass}, we estimate the SMBH masses and the Eddington ratios. Finally, we discuss the results in Section \ref{sec:discussion} and conclude in Section \ref{sec:conclusion}.

\section{The sample of isolated AGNs at z < 0.05 and X-ray data processing}
 \label{sec:sample}
\subsection{Sample}

Our study focuses on the sample of 61 isolated AGNs \citep{Chesnok2010, Pulatova2015}, which was obtained by cross-matching the 2MASS Isolated Galaxy catalogue (2MIG) by \cite{Karachentseva2010} with the Catalogue of quasars and AGNs by \cite{Veron2010}. The 2MIG criteria include limits for the visual magnitude $K_{s}$ $\leq$ 12.0 mag and velocity $V_{r}$ < 15,000 km/s of galaxies as well as the isolation parameters: the galaxies with the infrared angle diameters $a_{K}$ $\geq$ 30 arcsec have been considered as isolated ones if all their relevant neighbouring galaxies with angle diameters $a_{i}$ and angle distances $x_{i}$ satisfied the empirical conditions $x_{1i}/a_{i}$ $\leq$ 30 and 1/4 $\leq$ $a_{i}/a_{1}$ $\leq$ 4. This latter isolation criterion ensures that host galaxies have not undergone merging for at least 3 Gyr. The isolation criteria were verified many times in the AMIGA project framework (see, for example, an earlier work by \cite{2007Verley} for the Catalogue of Isolated Galaxies \citep{1973Karachentseva}) as well as in our works to check the closest environment of isolated AGNs \citep{Pulatova2015}, \citep{Kompaniiets2025}.

We searched the HEASARC database for available individual X-ray observations of isolated AGNs, covering the INTEGRAL, Chandra, XMM-Newton, NuSTAR, Suzaku, and Swift missions. Additionally, we cross-matched with the BAT AGN Spectroscopic Survey catalogue \citep{Ricci2017BAT}. We found only 26 of the 61 galaxies having available X-ray observations, of which only 13 are registered in the 14–195 keV energy range. Table \ref{tab:general} presents the X-ray fluxes in the soft and hard ranges for objects previously analysed by other research groups. For nine galaxies — NGC 6951, NGC 5231, NGC 1050, ESO 215-014, CGCG 179-005, and IC 5287 — X-ray analysis was performed for the first time as part of this work. Five galaxies (NGC 3035, UGC 12282, UGC 06769, UGC 10774 and UGC 02936) were excluded from further spectral analysis due to poor-quality data. Thus, the final sample of isolated active galactic nuclei with reliable X-ray observations consists of 21 objects.

\begin{table*}
\scriptsize
	\caption{Main parameters of the 2MIG isolated AGNs and values of their spectral flux densities in the X-ray range}
    \label{tab:general}
\begin{tabular}{|l|c|c|c|c|c|c|c|c|c|c|c|c}
\hline\hline
2MIG & Name  & RA, & Vh   & Morph. & Type &  F$_{obs, 2-10 keV}$  &   Data,   & F$_{obs, 14-195 keV}$    & Data, & 	Inclination \\

     &       & DEC & km/s & type   & AGN   &  $erg\ cm^{-2} s^{-1}$  &    2-10 keV    &  $erg\ cm^{-2} s^{-1}$   &   14-195 keV   &  angle, deg  \\
\hline 
   1 &         2         &      3 and 4  & 5    &  6           & 7       & 8 &   9         &10  & 11 & 12 \\
   \hline
320     &    NGC 1050      & 02:42:35.60  &  3699&	SBa  & Sy2& 2.08$^{+1.90}_{-2.03}$ $\cdot 10^{-14}$ & XMM-Newton & &  & 45.8  \\ 
     &                   &  34:45:48.72 &  &  &  &  & &  &   \\
 415 &  ESO 116-018      &  03:24:53.04  & 5397	&(R)SAB(r) & Sy2     & 3.02$^{+0.21}_{-0.33}$ $\cdot 10^{-13}$   &   &   &      & 90 \\  
     &                   & -60:44:18.1   &       &             &         &   &        &   &      &  \\
 417 & MCG-02-09-040     &  03:25:04.94  & 4495  & S0-a        & Sy2     &   &  &  1.10 $\cdot 10^{-11}$ & Swift & 74.9 \\  
  
     &                   & -12:18:28.5  &        &             &         &   &              &   &       & \\
 488 &  UGC 02936        &  04:02:48.25 & 3811	& SB(s)d      & Sy2     &  1.20 $\cdot 10^{-12}$ & Chandra & 1.10 $\cdot 10^{-11}$ & Swift  & 	80.6 \\   
     &                    &  01:57:56.6   &  &  &  &  & &  &  &  \\
1018 &  ESO 208-034     &  07:43:31.71  & 7587 & 	SBab pec: & Sy2  &  &  & 5.00 $\cdot 10^{-12}$ & INTEGRAL   & 	75.4 \\   
     &                    & -51 40 56.7   &  &  &  &  & &  &  (40-100 keV)  & \\
1086  & IC 2227 &  08:07:07.18  &  9687 & SBa  & Sy2 & 2.60$^{+0.14}_{-1.06}\cdot 10^{-13}$  & XMM-Newton & 2.23$^{+0.43}_{-0.74}\cdot 10^{-12}$ & NuSTAR & 	65.2 \\  
     &                   &  +36:14:00.53  &  &  &  &  & &  & (15-55 keV)   &   \\
1126  &    CGCG 179-005      &  08:25:10.23 &  	5698  & Sbc  & Sy1 &1.81$^{+0.06}_{-0.05}$  $\cdot 10^{-12}$ & XMM-Newton & &  &75.2   \\ 
     &                   &     37:59:19.78 &  &  &  &  & &  &   \\
1345 &  NGC 3035          &  09:51:55.02  & 4354 & SB(rs)bc & Sy 1  & 6.60 $\cdot 10^{-12}$ & Swift & 19.30 $\cdot 10^{-12}$  & Swift  & 	37.6 \\  
     &                   & -06 49 22.5   &  &  &  &  & &  &  & &   \\
1363 &  NGC 3081         &  09:59:29.53  & 2391  & (R)SAB(r)0/a & Sy2  &  2.50 $\cdot 10^{-12}$ & Swift & 83.20 $\cdot 10^{-12}$  & Swift   & 59.8 \\  
     &                   & -22:49:34.3   &  &  &  &  & &  &  &\\
1384 &  ESO 499-041        & 10:05:55.37 & 	3842 & (R)SB0(r) & AGN & 7.00 $\cdot 10^{-12}$ & Swift  &  14.60 $\cdot 10^{-12}$ & Swift  & 90.0 \\  
     &                   & -23 03 25.1    &  &  &  &  & &  &  &  \\
1442 &  ESO 317-038       &  10:29:45.61  & 	4542 & (R)SBa pec  & AGN & 4.00 $\cdot 10^{-13}$ & Swift & 12.10 $\cdot 10^{-12}$ & Swift  & 90.0 \\  
     &                   & -38 20 54.7   &  & &  &  & &  &    &  \\
1516   &  ESO 215-014        &  10:59:19.06  &  5963& SB(rs)b & Sy1&5.96$^{+0.85}_{-0.85}$  $\cdot 10^{-12}$ & Swift & &  &  60.4 \\ 
    &                   &    -51:26:33.160   &  &  &  &  & &  &   \\
1550 & ESO 438-009        &  11:10:47.97  & 	7198 & 	(R')SB(r)ab pec & Sy1 &  4.00 $\cdot 10^{-12}$ & Swift & 6.70 $\cdot 10^{-12}$  & Swift   & 50.8\\   
     &                   & -28:30:03.9   &  &  &  &  & &  &     & \\
1607 &  IGR J11366-6002 &   11 36 42.05 &  4467  & Sa   & Sy2/LINER &4.20 $\cdot 10^{-12}$ &  Swift & 20.70 $\cdot 10^{-12}$ & Swift & - \\  
     &                      &    -60 03 07.0 &  &  &  &  & &  &&   & \\
1633     &  UGC 06769        &  11:47:43.68   &  	8915 &  SB(r)b &  Sy 2 &9.00 $\cdot 10^{-13}$ &Swift & 19.20 $\cdot 10^{-12}$ &Swift  & 	67.0  \\   
     &        &    +01:49:34.420   &  &  &  &  & &  &   \\
1646 &  CGCG 243-024      &  11:53:41.76  & 	7270 & SBab & Sy1 &  1.26 $\cdot 10^{-12}$  & XMM-Newton &  &  & 31.2 \\  
     &                   &  46 12 42.6   &  &  &  &  & &  &   & \\
1873 &  NGC 5231         &  13:35:48.25  &6527   & SBa          & Sy1  & 6.40$^{+0.10}_{-0.20}\cdot 10^{-12}$  & XMM-Newton &  15.60 $\cdot 10^{-12}$ & Swift  & 53.0\\   
     &                   &  02:59:55.6   &  &  &  &  & &  &   &  \\
1915 &  NGC 5347         & 13:53:17.9  & 2368 & (R')SB(rs)ab &  Sy2  & 2.30 $\cdot 10^{-13}$  & Chandra &   &   & 45.3 \\  
     &                   &  33:29:26.7   &     &     &     &  &   &  &   &  \\
1950 &  ESO 097-013      &  14:13:09.9  & 434 & 	SA(s)b &      Sy2 &  1.27 $\cdot 10^{-11}$ & Swift & 272.10 $\cdot 10^{-12}$ & Swift & 64.3 \\  
     &                   & -65 20 20.4   &  &  &  &  & &  &   &  \\
2183 &  UGC 10120        &  15:59:09.7  & 9438 & 	SB(r)b     &   Sy1 &  3.50 $\cdot 10^{-12}$ & XMM-Newton &   &   & 55.7\\  
     &                   &  35:01:47.3   &  &  &  &  & &  &  &   \\
2357 &  UGC 10774        & 17:14:08.9  & 8931 & 	SABb    &   Liner &  8.39 $\cdot 10^{-15}$ & XMM-Newton &   &   & 60.6\\  
     &                   &  +58:49:06.2   &  &  &  &  & &  &  &   \\
2363 &  NGC 6300         &  17:16:59.47  & 1108 & 	SB(rs)b & Sy2 &  8.60 $\cdot 10^{-12}$ & Swift  & 99.40 $\cdot 10^{-12}$ & Swift  & 	52.7 \\  
     &                   & -62:49:13.9   &  &  &  &  & &  &   &   \\
2811 &  NGC 6951          &  20:37:14.07  & 1424 & 	SAB(rs)bc & Sy2 & \textbf{1.5$^{+0.4}_{-1.4} \cdot 10^{-13}$} & Swift  &  &  & 	50.8 \\  
     &                   &  66:06:20.3   &  &  &  &  & &  &    &  \\
3110 &  UGC 12282         &  22:58:55.28  & 	5094 &   Sa & Sy1.9 &  9.00 $\cdot 10^{-13}$ & Swift & 19.20 $\cdot 10^{-12}$ & Swift  & 90.0\\  
     &                   &  40:55:55.9   &  &  &  &  & &  &    & \\
3118 &  NGC 7479         &  23:04:56.7  &	2376  & SB(s)c & Sy2 &  3.00 $\cdot 10^{-13}$ & Swift  &  20.40 $\cdot 10^{-12}$ & Swift  & 43.0\\
     &                   &  12:19:22.3   &  &  &  &  & &  &    &  \\
3128 &  IC 5287          &  23:09:20.3  &	9713 &(R')SB(r)b & Sy1.2  & 3.90 $\cdot 10^{-13}$ & Chandra &  &    &  67.9\\
     &                   &  00:45:23.0   &  &  &  & (2-7 keV) & &  &  & & \\
     \hline
\end{tabular}
 \begin{minipage}{\linewidth}
\smallskip
{
NOTE: Column 1: number in the 2MIG catalogue. Column 2: Galaxy's name. Columns 3 and 4: equatorial coordinates (J2000.0). Column 5: radial velocity, in km/s, from the 2MIG catalogue. Column 6: morphological type. Column 7: spectral type of nuclear activity. Column 8: spectral flux densities, in $\rm{erg\ cm^{-2} s^{-1}}$, $E$=2--10 keV. Column 9: survey of space telescope for 2--10 keV range. Column 10: spectral flux densities, in $\rm{erg\ cm^{-2} s^{-1}}$, $E$=14--150 keV. Column 11: survey of space telescope for 14--150 keV range. Column 12: inclination angle between the line of sight and polar axis of a galaxy, in degrees from the HyperLEDA database \citep{Makarov2014A&A}. We used flux estimations for ESO 208-034 by \cite{2009Beckmann}, for UGC 10120 by \cite{2009A&ABianchi}, for ESO 116-018 by \cite{2019Marchesi}, for MCG-02-09-040 and UGC 02936 by \cite{2010A&ACusumano}, for UGC 10774 by \citep{Nisbet2016}. The Chandra fluxes for UGC 10120, UGC 02936, NGC 7479, NGC 6300, NGC 5347, and ESO 097-013 were taken from the article by \cite{2016Wang}. The SWIFT fluxes for NGC 3081, NGC 3035, ESO 438-009, UGC 06769, ESO 097-013, NGC 6300, UGC 12282, NGC 7479, ESO 317-038, ESO 499-041, ESO 243-024, and IGR J11366-6002 were adopted from the data by \cite{2017Ricci}. The observed fluxes for NGC 6951, NGC 5231, NGC 1050, ESO 215-014, CGCG 179-005, and IC 5287 are obtained in this work.
}
\end{minipage} 
\end{table*}

\subsection{X-ray observations and data reduction}

The \textbf{XMM-Newton} data were processed and cleaned using the Science Analysis Software (SAS version 19.0.0) with the appropriate calibration files. Each observation was processed with standard pipelines. Due to its larger effective area, only data from the EPIC/PN camera have been considered in this work, even if the data from EPIC/MOS1 and MOS2 were available. However, we did not use EPIC/PN data for CGCG 179-009 because the source spot lies at the gap between individual chips; therefore, we used observational data from two MOS cameras.

We extracted event files that contain only single- and double-pixel events. These files were filtered for FLAG = 0 quality and cleaned of high-background time intervals. The source counts were taken within circular regions with a 20 to 35 arcsec radius, depending on the source brightness. The background was extracted from similar areas. The offset was chosen close to the target source, avoiding the CCD edges, the out-of-time event strips, and other sources. Response matrices and auxiliary response files were created with \textit{arfgen} and \textit{rmfgen}.

The \textbf{Swift/XRT} data reduction was performed using the \textit{xrtpipeline} task, which is part of the XRT Data Analysis Software (XRTDAS) included in the HEAsoft 6.28 package. All data were extracted only in photon-counting (PC) mode, using the standard grade filtering 0-12 for PC. Source counts were extracted from a circular region with a radius of 15-25 arcsec, depending on the source's brightness. Background counts were extracted from a nearby, source-free circular region with an area approximately 3 times larger than that used to extract the source counts. We generate auxiliary response files using task \textit{xrtmkarf} and exposure maps with task \textit{xrtexpomap}, respectively. Due to a series of relatively short exposures for each observed source, we stacked the corresponding spectra together with standard FTOOLS routines \textit{(mathpha, addarf, addrmf)}. Spectral analysis was performed in the 0.3-10 keV band.

The time-averaged spectra obtained from XMM-Newton/PN and Swift/XRT data were grouped to a minimum of 20 counts per bin to facilitate the use of the $\chi^{2}$ minimisation technique in the spectral fitting. In cases of low-quality spectral data, we group the spectra so that each channel contains at least one count, and then perform spectral fitting using $C$-statistics.

The \textbf{Chandra} data reduction was performed with the \texttt{CIAO 4.13} software package \citep{2006Fruscione} and the calibration database {\tt CALDB\ 4.9.4}, after reprocessing the data with the \texttt{chandra\_repro} script, as recommended in the {\tt CIAO} analysis threads. Finally, we extracted the source and background spectra for Chandra imaging mode with the \textit{specextract CIAO} script. We note that archival observational data for PGC 35009 and UGC 06087 were not processed due to their low quality. 

The \textbf{NuSTAR} data reduction was performed with NuSTARDAS 2.0.0 using the CALDB v20200813 calibration files and the standard settings of the \textit{nupipeline} task for each of the FPMA and FPMB detectors. We extracted source counts in circular regions with a radius of 40-50 arcsec, as well as the background counts from the region between two circles centred on the source position with an inner radius of 80-90 arcsec and an outer radius of 120-130 arcsec. Spectra were processed with the standard \textit{nuproducts} tool. The data reduction for NGC 5347 and MCG-02-09-040 is described in our paper by \cite{Vasylenko2020}.

\section{Spectral models of isolated AGNs}
\label{sec:Xray_generall}

We employed various spectral models to investigate the specified AGN parameters of isolated AGNs. We used XSPEC version 12.11.0 to perform spectral modelling and analysis of EPIC/PN and XRT data. The spectra for faint sources were rebinned to ensure a minimum of 3 counts per energy bin and use the C-statistic (the ‘W-statistic’ in XSPEC). Additionally, we performed spectral analysis of the Chandra data using Sherpa (ver. 4.12), the CIAO modelling and fitting application. The lower and upper limits as well as the errors correspond to a 90 $\%$ confidence level for the single parameter (i.e., 1$\sigma$ uncertainties).

To consider the individual characteristics of each isolated AGN, we exploited different models to describe their spectra. Here we present the best results of model combinations. Such a source-specific approach was chosen over a unified modelling scheme to account for the diversity in spectral characteristics across our sample, particularly for objects exhibiting moderate or high absorption. Applying a single, simplified model to all sources could introduce substantial model dependence, potentially distorting estimates of intrinsic luminosity and spectral indices. Additionally, a uniform modelling framework may provide inadequate spectral descriptions of complex sources, such as Compton-thick nuclei. So, we performed spectral analysis of each galaxy according to its AGN type and at a level of detail permitted by the individual data quality, to ensure the reliability of the resulting parameter dependencies. For all spectral models, we used the Galaxy absorption ($N_{H, Gal}$) with \textit{Tbabs} model \citep{2000ApJWilms}. The value of $N_{H, Gal}$ for each galaxy was determined by the HEASARC tool\footnote{https://heasarc.gsfc.nasa.gov/cgi-bin/Tools/w3nh/w3nh.pl} of the \textit{nH FTOOL} \citep{1999Blackburn} with the HI4PI Map \citep{2016A&AHIPI}.

\subsection{NGC 6951}
\label{NGC6951}

The \textbf{NGC 6951} is classified as an SAB(rs)bc galaxy at $z$ = 0.00475 \citep{1998AJHaynes} with a Sy2-type nucleus \citep{2006Veron}. The X-ray spectrum of NGC 6951 obtained by SWIFT/XRT\footnote{Because this source is very faint in the x-ray range, we used all published observational data of Swift/XRT between 2015$\div$2024 years} is fitting by the model comprised of redshifted black body emission (\textit{zbbody}) and power law (\textit{$zpowerlw$}) for describing primary emission. The initial model included only the power law component. Because a spectrum relative to the model has appeared clearly "distorted", the black body model was also included despite good statistics (cstat/d.o.f = 26/22). However, this new approach (cstat/d.o.f = 17/22) required a non-physically flat slope of $\Gamma \approx 1$ with a clearly visible lowering of the data point near 2 keV. To resolve this, we included a simple neutral absorption component \textit{zTBabs}. In the XSPEC formalism, this final model is represented as follows: 

\begin{center}
Model A: $TBabs(zbbody + zTbabs \times zpowerlw)$
\end{center}

\begin{figure}
\centering
\rotatebox[origin=c]{0}{
\includegraphics[width=0.90\linewidth]{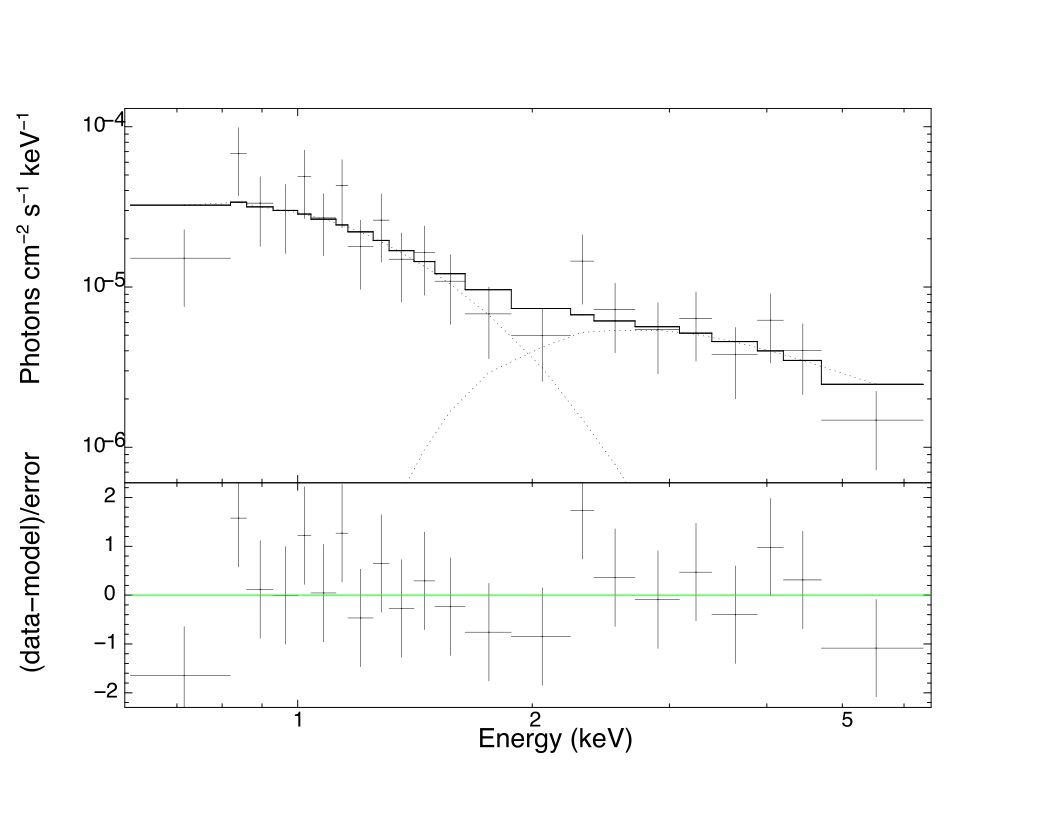}%
}
\vspace*{-7mm}
\caption{The NGC 6951 SWIFT/XRT spectrum fitted by model A.}
\label{fig:NGC6951}
\end{figure}

It provides good statistics (cstat/d.o.f = 15/28) with the best-fit values of the photon index $\Gamma = 2.13^{+0.83}_{-0.88}$ (Table \ref{tab:X-rayParam}), black-body temperature of $kT_{e} = 0.27^{+0.07}_{-0.05}$ keV and N$_{H} = 4.0^{+2.7}_{-1.9}\cdot 10^{22}$ cm$^{-2}$. The resulting spectrum of NGC 6951 is shown in Figure \ref{fig:NGC6951}.

\subsection{ESO 317-038}
\label{ESO317038}
The \textbf{ESO 317-038} is an (R)SBa galaxy at $z$ = 0.01515 \citep{2003A&APaturel} with a Sy2 candidate Compton-thick AGN \citep{Vavilova2015}. Here we present its spectral analysis with Swift/XRT and NuSTAR data. The choice of model for the spectral analysis is based on articles by \cite{2016ApJKoss} and \cite{2017Ricci}. Specifically, we used model \textit{B1} from the article by \cite{2017Ricci} (see Table 4 therein) as a basis. The difference in our application is that a) instead of the \textit{pexrav} model, we used the \textit{pexmon} model, which self-consistently includes the Fe $K_{\alpha}$ emission line, and b) it turned out that adding the \textit{cabs} (i.e., optically thin Compton scattering) component to the model did not improve the statistics at all. Thus, our model includes three components in the X-ray spectrum: the transmitted primary emission as an absorbed power-law \textit{$zTBabs \times zpowerlw$}, scattered unabsorbed emission \textit{$constant\times zpowerlw$}, and a neutral Compton reflection component with the \textit{pexmon} model. We have added a calibration constant between the NuSTAR and SWIFT/XRT spectral data to account for differences in normalisation across instruments and long-time spectral variations between observations. In the XSPEC formalism, the final model is presented as follows: 

\begin{center}
Model B: $TBabs(constant\times zpowerlw 
+ zTBabs \times zpowerlw + constant\times pexmon)constant $
\end{center}

\begin{figure}
\centering
\rotatebox[origin=c]{-90}{%
\includegraphics[width=0.6\linewidth,trim=20 20 0 0, clip]{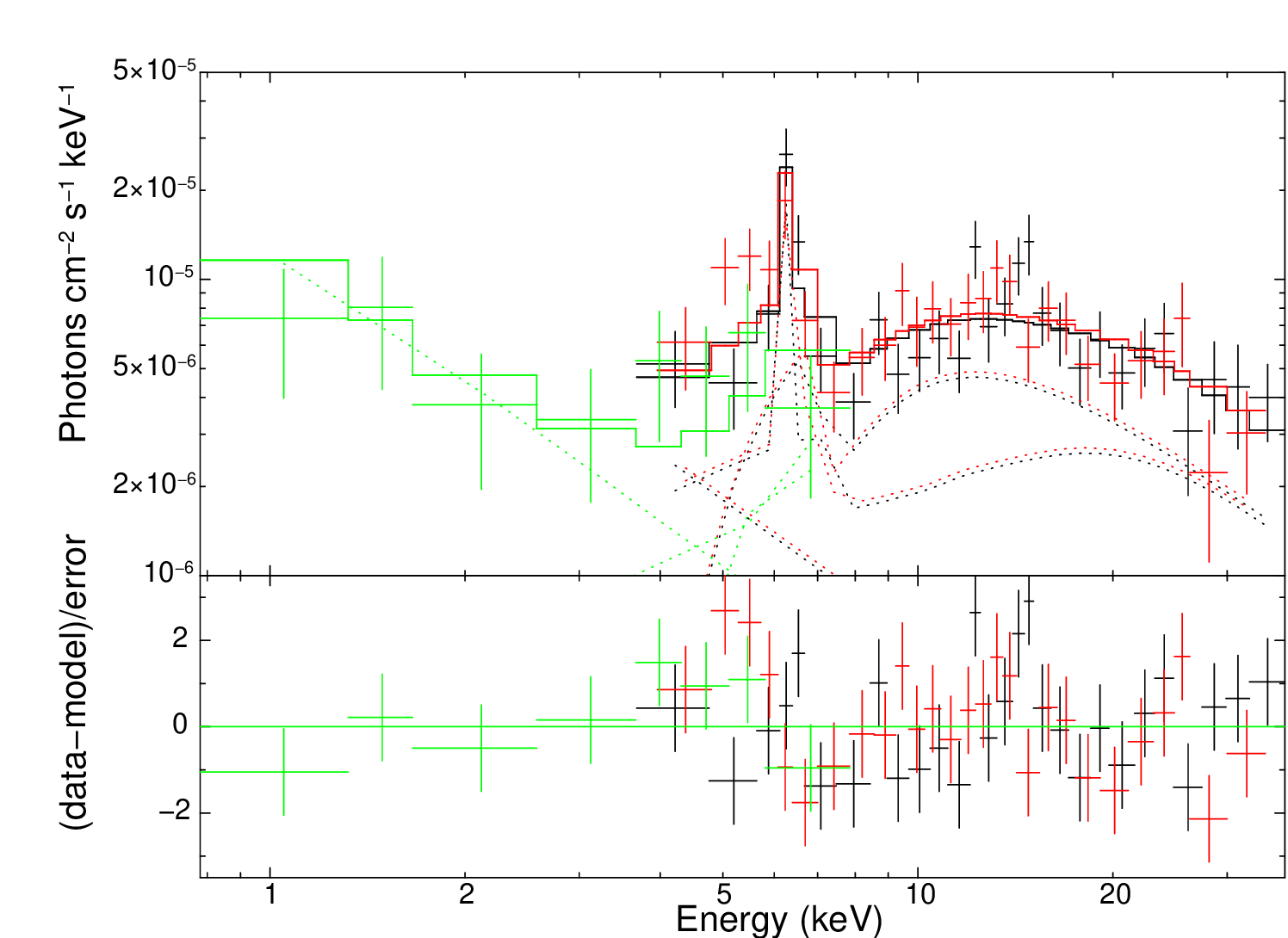}}
\vspace*{-12mm}
\caption{The ESO 317-038 NuSTAR and SWIFT/XRT joint spectrum fitted by model B.}
\label{fig:ESO317-038}
\end{figure}

This model yields good statistics (cstat/d.o.f. = 68/56) with best-fit values for the photon index, neutral absorption, and unabsorbed luminosity (Table \ref{tab:X-rayParam}). Figure \ref{fig:ESO317-038} represents the best-fit model for the X-ray spectrum of ESO 317-038.

\subsection{NGC 3081}
\label{NGC3081}
The \textbf{NGC 3081} is the S0/a multi-ringed galaxy at $z$ = 0.007976 \citep{2005A&ATheureau} with a Sy2 Compton-thin nucleus (but close to the Compton-thick threshold with $N_{H,l.o.s.} \sim 6 \cdot 10^{23} cm^{-2}$) \citep{Traina2021}. Our spectral study of NGC 3081 with Chandra observations was described in detail by \cite{Kompaniiets2023}. The spectral model, which we used, includes the following components (in Sherpa CIAO formalism): 

\begin{center}
Model C: $xstbabs (xsapec + xsapec + xsztbabs \times xszpowerlw + xszgauss)$
\end{center}

\begin{figure}
    \centering
    \includegraphics[width=0.95\linewidth]{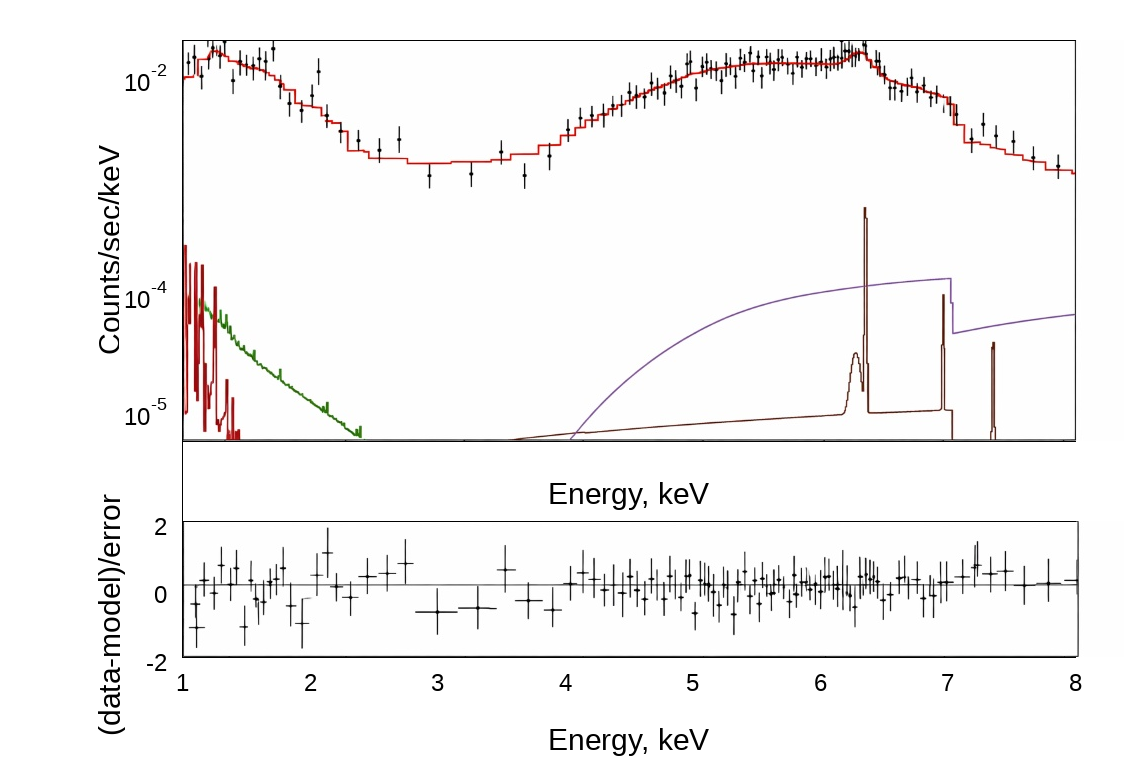}
    \caption{The NGC 3081 Chandra spectrum fitted by model C.}
    \label{fig:NGC3081}
\end{figure}

It consists of a power-law continuum (\textit{xszpowerlaw}), additional neutral absorption (\textit{xsztbabs}), an emission line with a Gaussian profile (\textit{xszgauss}), and two thermal components of diffuse gas in collisional ionisation equilibrium (\textit{xsapec}). Galactic absorption was fixed with value $N_{H} = 0.04 \cdot 10^{22} cm^{-2}$. This model provides good statistics with $\chi^{2}\d.o.f. = 99.76/103$. The soft excess is well fitted with two thermal components, where the temperature of the second component $kT_{2} = 1.0_{-0.1}^{+3.0}$ keV is much higher ("hot" component) in comparison to the first one of $kT_{1} = 0.16_{-0.02}^{+0.10}$ keV (''warm" component). The Fe $K_{\alpha}$ emission line $E_{line} = 6.39_{-0.02}^{+0.06} \rm{keV}$ has an equivalent width of $\textit{EW}$ $= 50_{-0.01}^{+0.01}$ eV. This suggests that the Fe $K_{\alpha}$ emission line originated in the neutral moderate-density environment. The spectral model is shown in Figure \ref{fig:NGC3081}.

\subsection{IC 5287}
\label{IC5287}
\begin{figure}
    \centering
\includegraphics[width=0.95\linewidth]{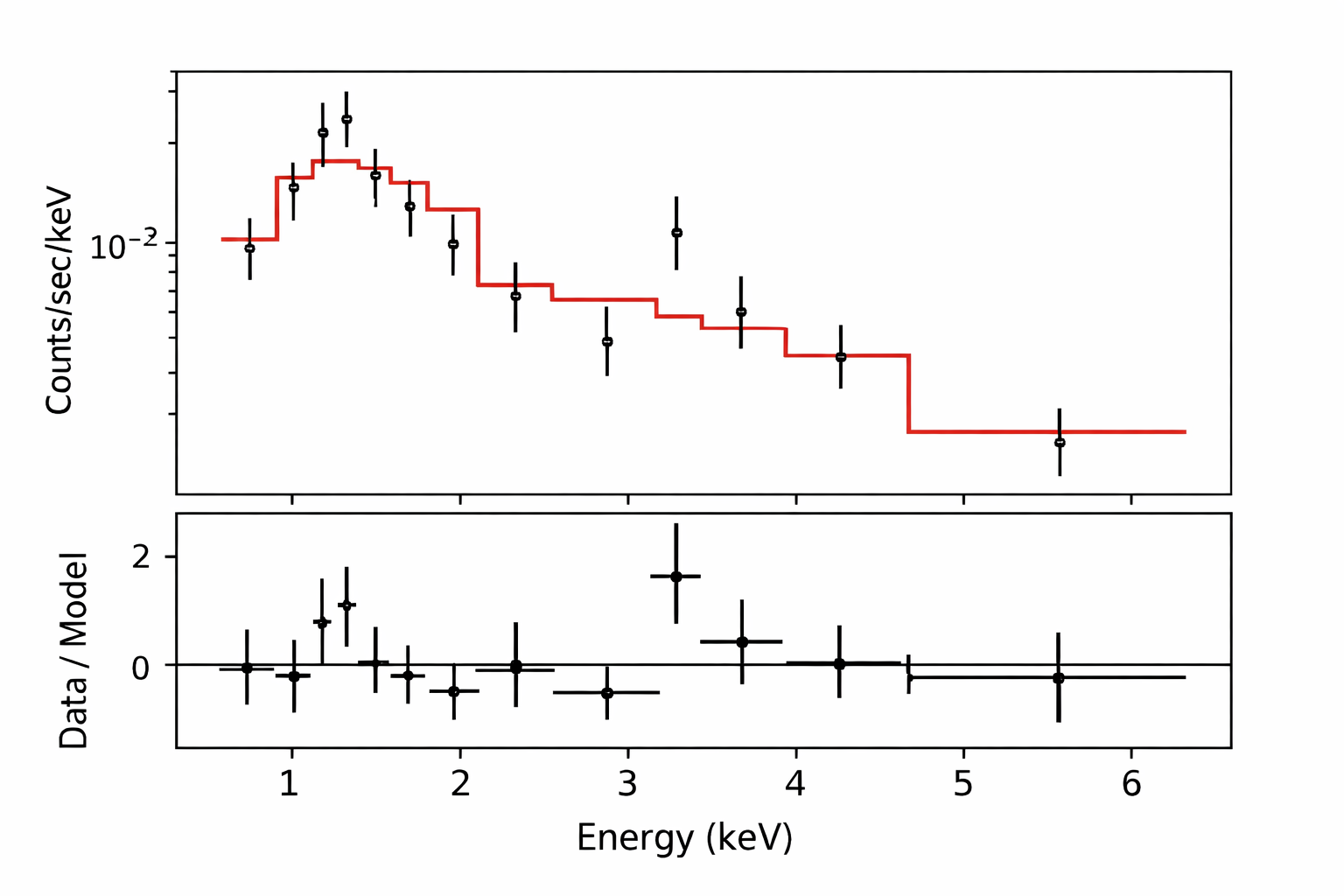}
    \caption{The IC 5287 Chandra spectrum fitted by Model D.}
    \label{fig:IC5287}
\end{figure}

The \textbf{IC 5287} is the (R')SB(r)b galaxy at z = 0.032399 \citep{2018PASPSDSS} with Sy 1.2 nucleus \citep{2006Veron}. For the fitting of the Chandra data, we used a simple absorbed power-law model \textit{$zTBabs\times zpowerlw$} (see Figure \ref{fig:IC5287}). 

\begin{center}
Model D: $TBabs(zTBabs\times zpowerlw)$
\end{center}

Its fit quality is moderate with cstat/d.o.f. = 16.25/11 with the best-fit values of photon index, neutral absorption, and unabsorbed luminosity which are given in the Table \ref{tab:X-rayParam}). 

\subsection{CGCG 243-024 (Mrk 42)}
\label{CGCG243024}

The \textbf{CGCG 243-024 (Mrk 42)} is classified as an SBa galaxy at a redshift of z = 0.0246 \citep{Falco_1999} and contains a Narrow-Line Sy 1 nucleus \citep{Malkan_1998}. To start the analysis of its X-ray spectra from XMM-Newton observation, we apply the basic power-law model to the part of the spectrum in the >2 keV range, which showed good results with $\Gamma=1.85\pm0.17$ and $\chi^{2}$/d.o.f.=24/25. Taking into account data at energies up to 0.3 keV clearly shows the presence of a soft excess ($\chi^{2}$/d.o.f. = 2323/203). Consequently, we added a blackbody emission model \textit{zbbody}, which significantly improved the statistics ($\chi^{2}$/d.o.f. = 240/201). However, the behaviour of the spectrum at energies $\lesssim$2.5 keV clearly showed features of an unaccounted component with emission lines, so we added the \textit{mekal} model – emission component from hot diffuse gas ($\Delta \chi^{2}$ = -49 for 3 d.o.f.). Also, for clarity, we replaced the simple \textit{zpowerlaw} with the better \textit{nthComp} component to describe the continuum. In the XSPEC formalism, this final model can be represented as follows: 

\begin{center}
 Model E: $TBabs(zbbody + mekal + nthComp)$   
\end{center}

\begin{figure}
\centering
\rotatebox[origin=c]{-90}{%
\includegraphics[width=0.6\linewidth,trim=20 20 0 0, clip]{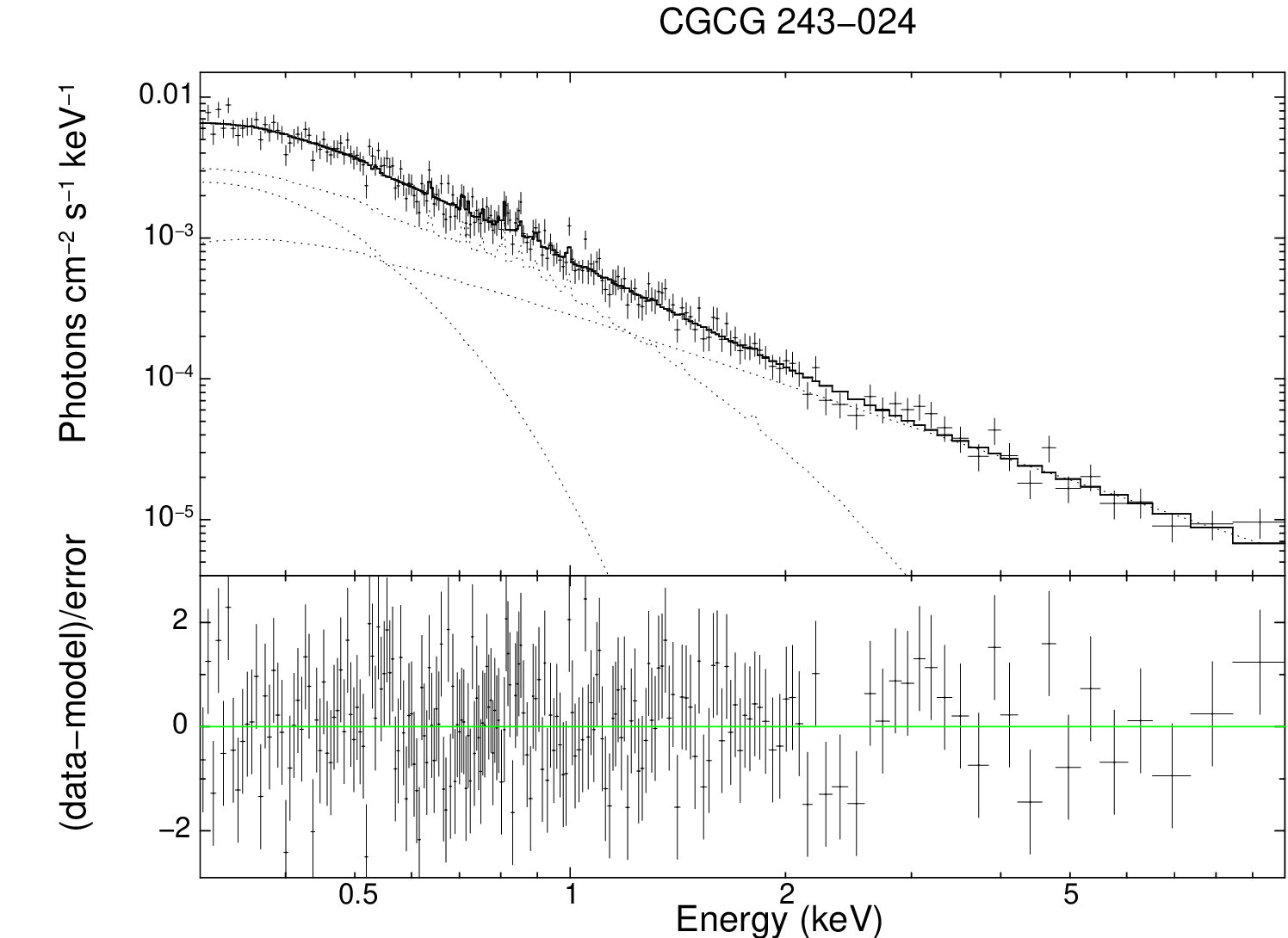}%
}
\vspace*{-8mm}
\caption{The CGCG 243-024 XMM-Newton spectrum fitted by model E.}
\label{fig:CGCG243-024}
\end{figure}

The best-fit values ($\chi^{2}$/d.o.f. = 192/198) of photon index is $\Gamma = 1.76^{+0.07}_{-0.02}$, and corresponding unabsorbed luminosity is L$_{2-10 keV}= 1.69^{+0.12}_{-0.13} \cdot 10^{42} \rm{ergs/s}$. Figure \ref{fig:CGCG243-024} represents the best-fit model for the X-ray spectrum of CGCG 243-0242.

\subsection{NGC 5231}
\label{NGC5231}
 
The \textbf{NGC 5231} is the SBa galaxy at z = 0.02177 \citep{2017SDSS} with a Sy 1 nucleus \citep{2006Veron}. Initially, we applied a simple power-law model over the 2-10 keV energy range. It showed a bad fit with $\chi^{2}$/d.o.f.=502/319 and a very flat spectral index $\Gamma =1.11\pm0.04$. Therefore, we added a neutral absorption component \textit{zTBabs} and obtained a good fit with $\chi^{2}$/d.o.f. = 306/318 with a spectral index $\Gamma =1.82\pm0.10$ and N$_{H}\approx2.32 \cdot 10^{22}$ cm$^{-2}$. When we included spectral data between 0.5 and 2.0 keV in the analysis with the current model, we observed a "soft" emission component ($\chi^{2}$/d.o.f. = 786/404). To describe this feature, we applied the \textit{apec} model (soft emission from collisionally ionised diffuse gas). This gave us very good statistics of $\chi^{2}$/d.o.f.=408/401, the temperature of the soft component kT$_{e}=0.86^{+1.23}_{-0.16}$ keV and $\Gamma=1.63\pm0.06$. Then we included the Swift/BAT spectrum (up to 75 keV) in the analysis, using a calibration constant to relate the XMM-Newton and Swift/BAT spectral data and account for differences in normalisation across instruments and long-term spectral variations between observations. The current model maintains a good fit statistic $\chi^{2}/$ d.o.f. = 414/407. But the behaviour of the Swift/BAT spectrum is ‘bumpy’; thus, to account for a possible so-called Compton hump, we included a non-relativistic reflection component \textit{xillver}. In the XSPEC formalism, the final model is presented as follows: 

\begin{center}
Model F: $TBabs(apec + zTBabs \times (zpowerlw + constant \times xillver))\times constant$
\end{center}

\begin{figure}
     \rotatebox[origin=c]{0}{\includegraphics[width=0.95\linewidth]{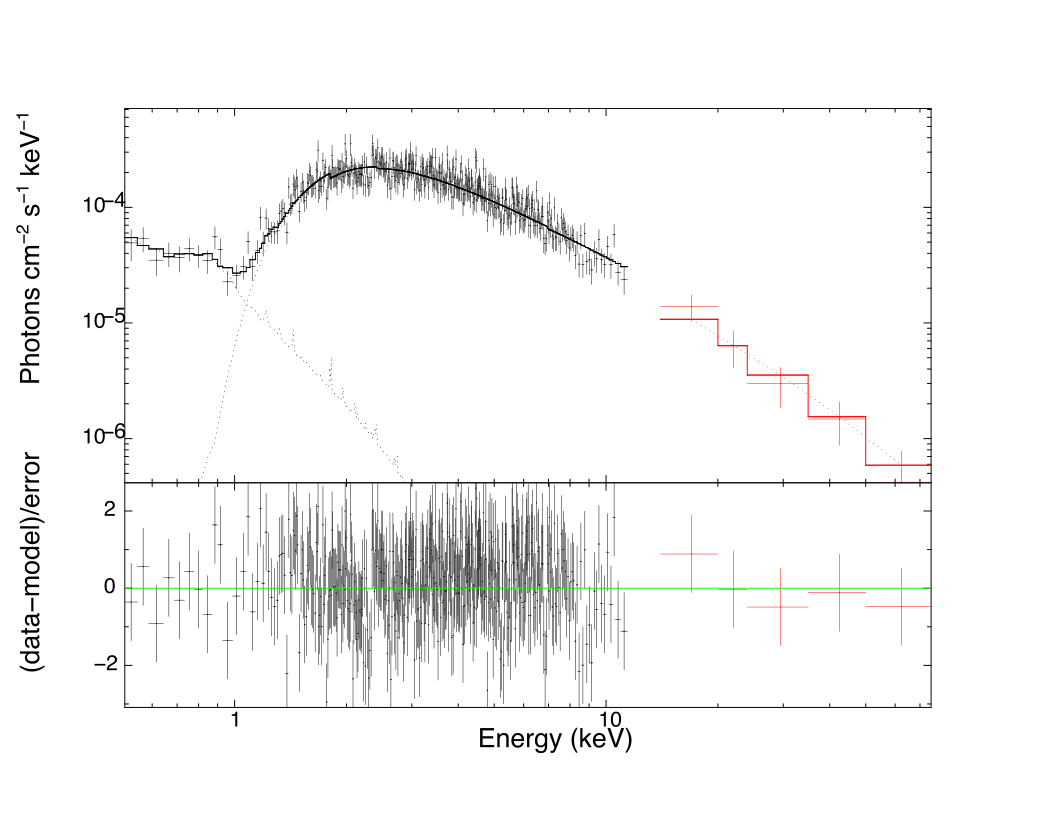}} 
 \caption{The NGC 5231 (black points) and the SWIFT/XRT (red points) joint spectrum fitted by Model F.}
   \label{fig:NGC5231}
\end{figure}

This model fits well with the final statistics of $\chi^{2}$ / d.o.f. = 408/403. We obtained the following best-fit values of parameters: $\Gamma = 1.67^{+0.05}_{-0.04}$, N$_{H} = 2.30^{+0.13}_{-0.13} \cdot 10^{22}$ cm$^{-2}$, ionisation parameter of $log\xi = 4.42_{-0.01}^{+0.01}$, and the temperature of the soft component $kT_{e} = 0.86_{-0.16}^{+0.59}$ keV. The corresponding unabsorbed model luminosity is L$_{2-10 keV}= 8.13^{+0.38}_{-0.37} \cdot 10^{42} \rm{ergs/s}$. The X-ray spectrum of NGC 5231 is shown in Figure \ref{fig:NGC5231}.

\subsection{CGCG 179-005}
\label{CGCG179005}

\begin{figure}
     \rotatebox[origin=c]{-90}{\includegraphics[width=0.60\linewidth,trim=20 20 0 0, clip]{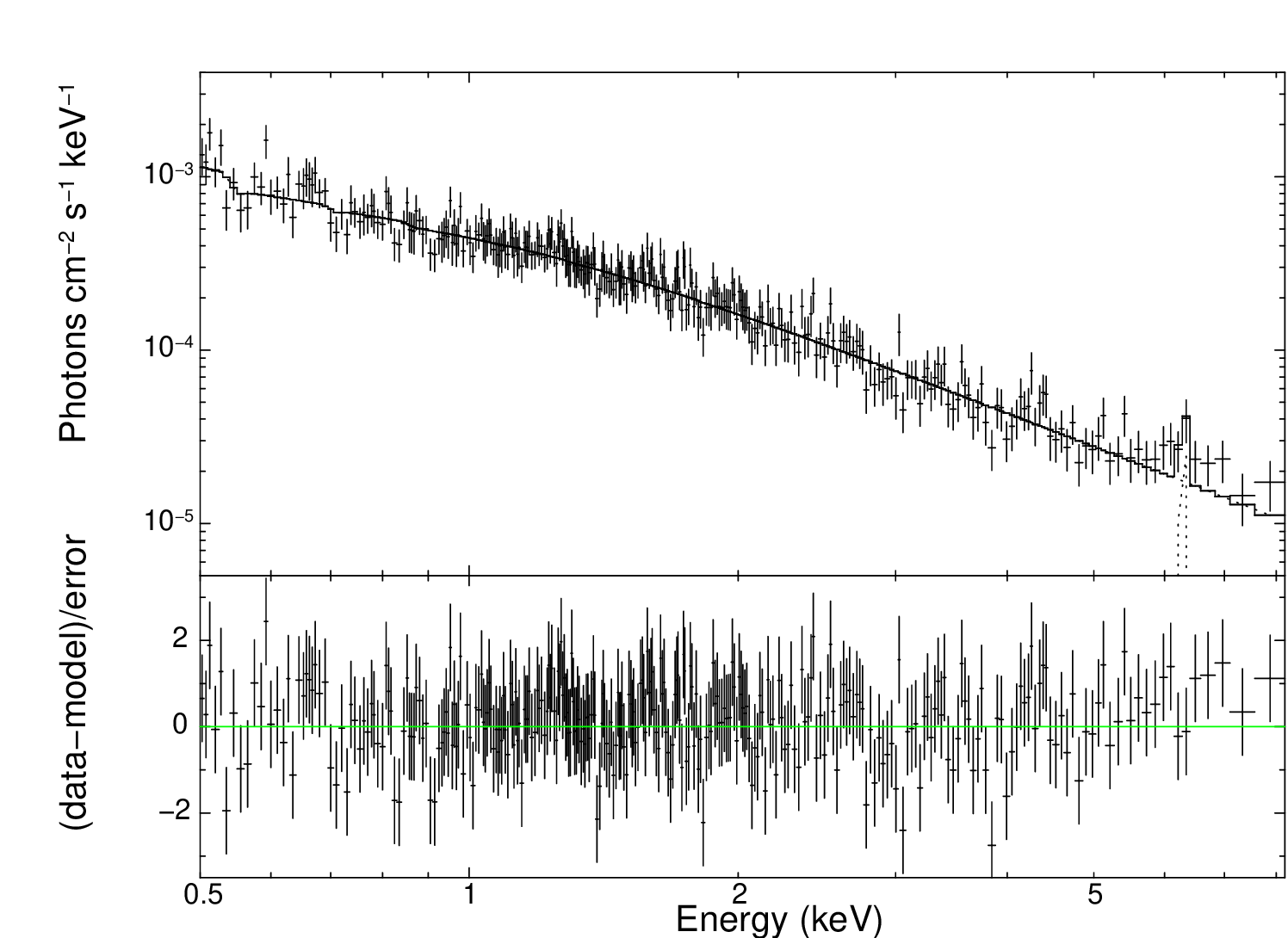}}
 \vspace*{-12mm}
 \caption{The CGCG 179-005 XMM-Newton spectrum fitted by Model G. }
   \label{fig:CGCG179-005}
\end{figure}

The \textbf{CGCG 179-005} is the Sc galaxy at $z$ = 0.021365 \citep{2017SDSS}, with the broad-line Sy 1 nucleus \citep{Liuplus2019}. We started the analysis of the X-ray spectrum from the merged XMM-Newton/MOS1+MOS2 data in the 0.5-8 keV energy range by fitting with a simple power-law model, which shows good statistics ($\chi{^2}$/d.o.f.). = 279/315 and a power index of $\Gamma=1.70\pm0.03$. Despite the good statistics, the spectrum shows slightly distorted behaviour in the 1$\div$3 keV range. Accordingly, we added a simple neutral absorption component \textit{zTBabs}, which showed a value of N$_{H} = 6.0\pm2.3\cdot 10^{20}$ cm$^{-2}$, $\Gamma=1.86\pm0.07$ with statistics of $\chi^{2}$ /d.o.f. = 259/314. However, the shape of the spectrum still remains somewhat inconsistent at energies $\lesssim$1.5 keV. Therefore, we replaced the \textit{zTBabs} component with a partial neutral absorption model \textit{zTBpcf}. The obtained spectral parameters are: $\Gamma=1.98\pm0.12$, N$_{H} =3.8\pm2.2\cdot 10^{21}$ cm$^{-2}$ ($\chi^{2}$ /d.o.f. = 253/313). The visible emission at 6 keV was modelled by a narrow Gaussian and can be interpreted as a narrow Fe K$_{\alpha}$ emission line. Hence, the final model in XSPEC is (see  Figure \ref{fig:CGCG179-005}):

\begin{center}
Model G: $TBabs(TBpcf \times zpowerlw + zgauss)$
\end{center}

The best-fit parameters with a fit statistic of $\chi^{2}$/d.o.f. = 247/317 are given in Table \ref{tab:X-rayParam}. An emission narrow Fe K$_{\alpha}$ line centroid energy is determined as $E_{line}=6.42_{-0.17}^{+0.28}$ keV with an equivalent width of $EW=219_{-121}^{+144}$ eV. The unabsorbed luminosity is L$_{2-10 keV} = 2.19^{+0.50}_{-0.10} \cdot 10^{42} \rm{erg/s}$. 

\subsection{ESO 215-014}
\label{ESO215014}
The \textbf{ESO 215-014} is the SB(rs)b galaxy at z = 0.01901 \citep{1991Vaucouleurs} with a broad-line Sy 1 nucleus \citep{Veron2010}. The Swift/XRT spectrum was fitted by Model H, which contains only a power-law continuum and Galactic absorption, i.e., in the XSPEC formalism: 

\begin{center}
Model H: ${TBabs \times zpo}$
\end{center}

\begin{figure}
     \rotatebox[origin=c]{-90}{\includegraphics[width=0.6\linewidth,trim=20 20 0 0, clip]{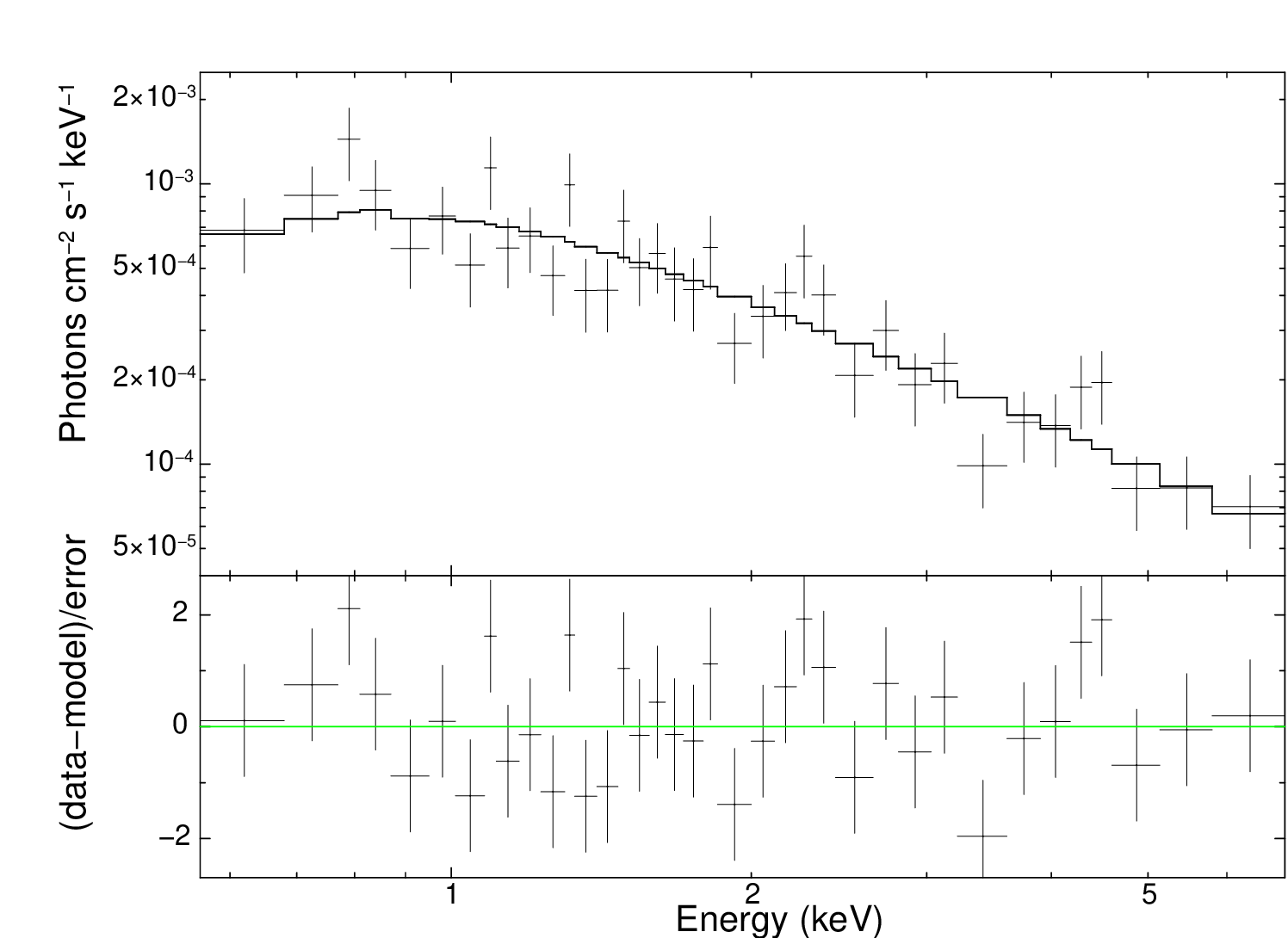}}
\vspace*{-12mm}
 \caption{The ESO 215-014 XMM-Newton spectrum fitted by Model H. }
   \label{fig:ESO215-014}
\end{figure}

As a result, we obtained a best-fit statistic of cstat/d.o.f = 37/35 (Figure \ref{fig:ESO215-014}). The corresponding values of the photon index are $\Gamma = 1.58^{+0.14}_{-0.13}$ and the unabsorbed luminosity in the 2-10 keV is L$_{2-10 keV}$ $ 4.89^{+0.77}_{-0.67} \cdot 10^{42}$ \rm{ergs/s}. 

\subsection{NGC 1050}
\label{NGC1050}
The \textbf{NGC 1050} is the SBa galaxy at $z$ = 0.01292 \citep{2017SDSS}, with the Sy 2 nucleus \citep{Veron2010}. Its spectral model was previously explored by \cite{Vavilova2015}. We began analysing the X-ray spectrum of the XMM-Newton/PN observation using a simple $tbabs\times zpo$ model, which yielded poor results ($\Gamma=2.85\pm0.23$, cstat/d.o.f.=29/17) with significant deviations at energies $\lesssim$1.5 keV. Therefore, we included an additional blackbody component \textit{zbbody}. This significantly improved the fit, as shown by the cstat/d.o.f. = 16/15 statistics and $kT_{e}=0.19\pm0.02$ keV. However, at energies $\lesssim$2 keV, the spectrum appeared "bumpy", so we also added neutral absorption \textit{zTBabs} and a component \textit{cabs} to account for optically thin scattering. We also tried applying the partial absorption component \textit{TBpcf} instead of \textit{cabs$\times$ ztbabs}, but in this case, the covering factor was not determined. In the XSPEC formalism, the final variant of the model is presented as follows:

\begin{center}
Model I: ${TBabs(zbbody + zTBabs \times cabs \times zpo)}$
\end{center}
\begin{figure}
     \rotatebox[origin=c]{-90}{\includegraphics[width=0.60\linewidth,trim=20 20 0 0, clip]{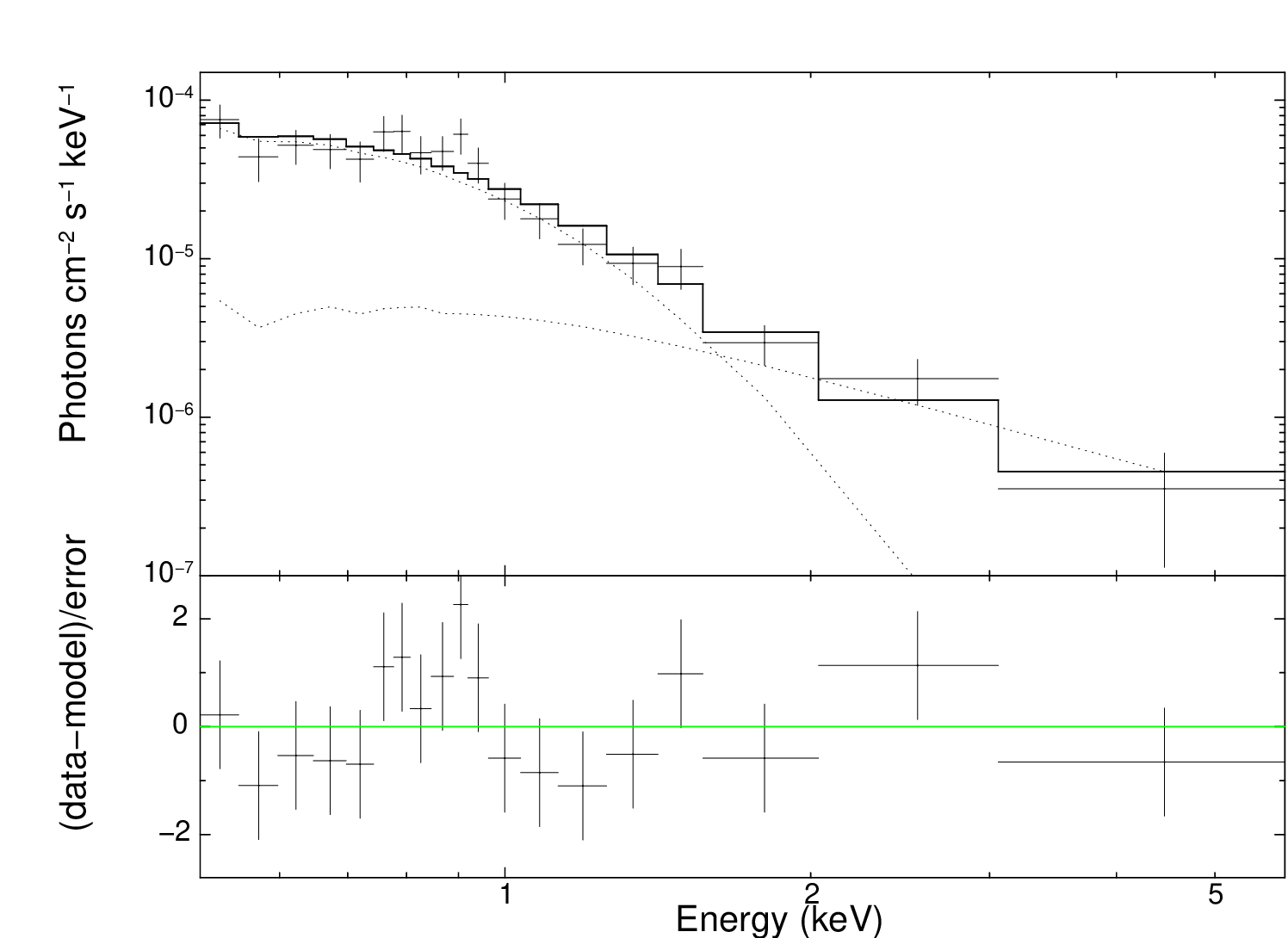}}
\vspace*{-12mm}
 \caption{The NGC 1050 XMM-Newton spectrum fitted by Model I. }
   \label{fig:NGC1050}
\end{figure}
As a result, we obtained the best-fit statistics of cstat/d.o.f = 15/15 (Figure \ref{fig:NGC1050}). The fit returns a value of the photon index of $\Gamma = 1.99^{+1.82}_{-1.18}$ and black-body temperature of $kT_{e} = 0.19\pm{0.02}$ keV. The corresponding unabsorbed model luminosity in 2-10 keV is L$_{2-10 keV}$ = $7.76^{+2.78}_{-4.38} \cdot 10^{39}$ \rm{ergs/s}. 

\subsection{NGC 5347, ESO 438-009, MCG-02-09-040, IGR J11366-6002}
\label{othersbyVasylenko}

In our previous work by \cite{Vasylenko2020}, we have provided the spectral models for the \textbf{NGC 5347}, \textbf{ESO 438-009} (see also \cite{Vavilova2015}), \textbf{MCG-02-09-040}, \textbf{IGR J11366-6002} galaxies with NuSTAR data. The Swift/BAT data up to $\approx$150 keV for MCG 02‑09‑040, ESO 438‑009, and IGR J11366‑6002, as well as the Swift/XRT data in the 0.3–10 keV energy band for NGC 5347, ESO 438‑009, and IGR J11366‑6002, were explored. The spectra of two sources, NGC 5347 and MCG‑02‑09‑040, show the Fe K$\alpha$ emission line with a significant EW $\approx$ 1 keV value. The X‑ray spectrum of NGC 5347 is best fitted by a pure reflection model with $E_{cut}$$\approx$ 117 keV and without any additional primary power‑law component. The X‑ray spectrum of MCG‑02‑09‑040 shows the presence of heavy neutral obscuration, and its fit is obtained by adopting the physical Monte Carlo‑based model as BNTorus. The spectral parameters and models of these isolated AGNs are included in Table \ref{tab:X-rayParam}. 

\subsection{ESO 499-041: \\
\textbf{a tentative relativistic interpretation of the Fe K feature}}
\label{ESO499041}

\textbf{ESO 499-041} is an (R1)SB:(r)0 galaxy at $z$ = 0.012816 \citep{2012ApJS2MASS} with a narrow-line Sy 1 nucleus \citep{2016Schmidt}. This source exhibits relatively high long-term variability in the flux amplitude, with a factor of $\sim$2.5 observed by Swift/XRT. To analyse its X-ray spectrum, we used a model comprising a soft component describing a redshifted thermal bremsstrahlung component, \textit{zbremss}, and a primary continuum component absorbed by a power law, \textit{$zTBabs \times zpo$}. In the XSPEC formalism, this model can be represented as follows:

\begin{center}
Model J: $TBabs (zbremss + zTBabs \times zpo)$ 
\end{center}

\begin{figure}
\centering
\adjustbox{valign=c,center}{\rotatebox[origin=c]{-90}{\includegraphics[height=0.8\columnwidth,trim=20 20 0 0, clip]{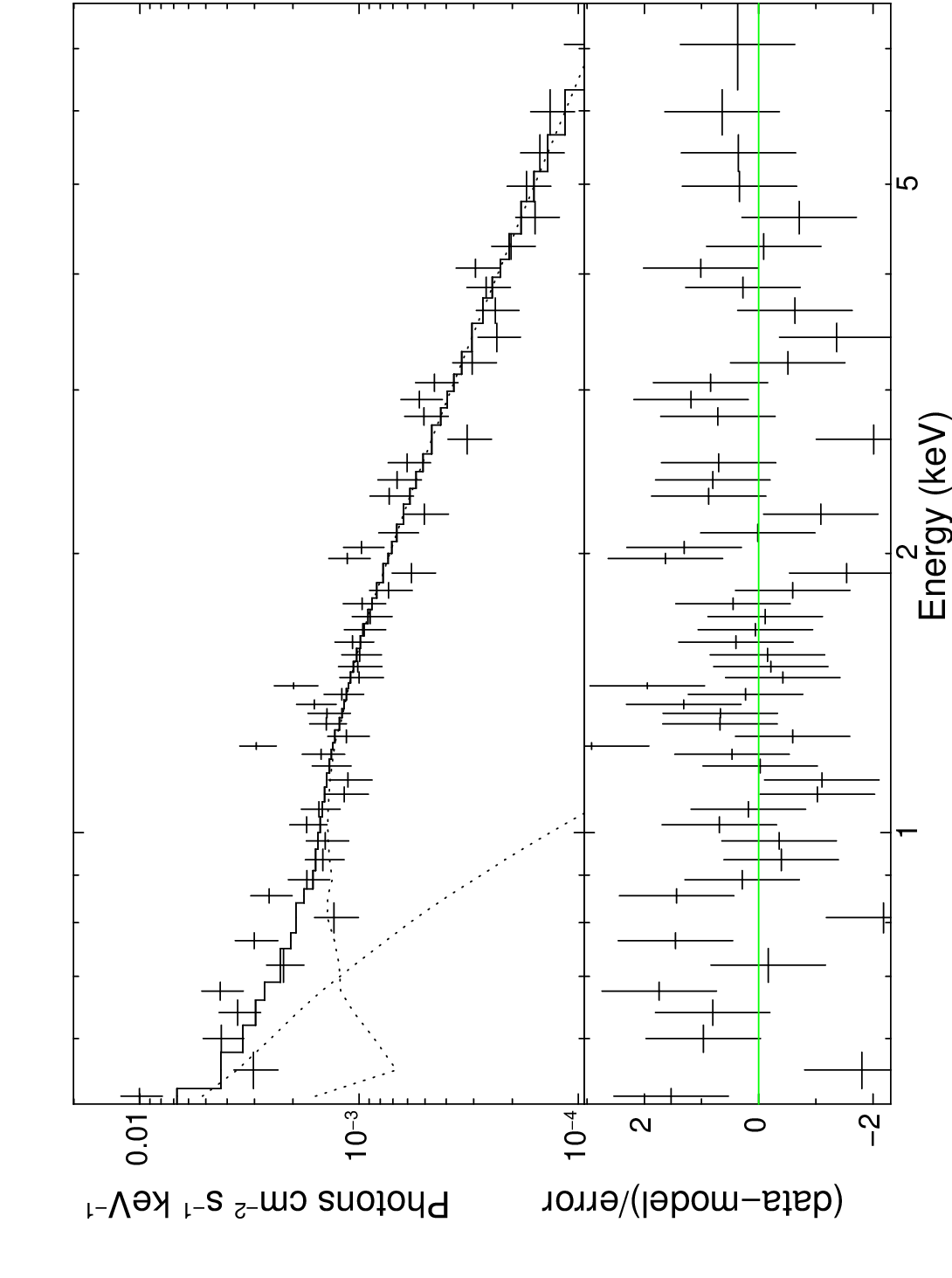}}}
\caption{The ESO 499-041 best-fit Model J for the SWIFT/XRT spectrum. The upper panel shows the best-fit model overlaid on the unfolded spectrum; the residuals from the fit are shown in the bottom panel.}
\label{fig:ESO499Swift}
\end{figure}

Figure \ref{fig:ESO499Swift} represents the best-fit Model J with final statistics of $\chi^{2}$/d.o.f. = 60/56. The best-fit photon index and absorption values are given in Table \ref{tab:X-rayParam}. We also found that the plasma temperature for the soft component is $kT_{e} = 0.16_{-0.08}^{+0.69}$ keV. The unabsorbed luminosity of this analysed bright state $L_{2\text{--}10\,\mathrm{keV}}$ is \\
$3.72^{+0.44}_{-0.41} \times 10^{42}\,\mathrm{erg\,s^{-1}}$. \\[1ex]
\begin{figure}
    \centering
    \includegraphics[width=0.85\linewidth]{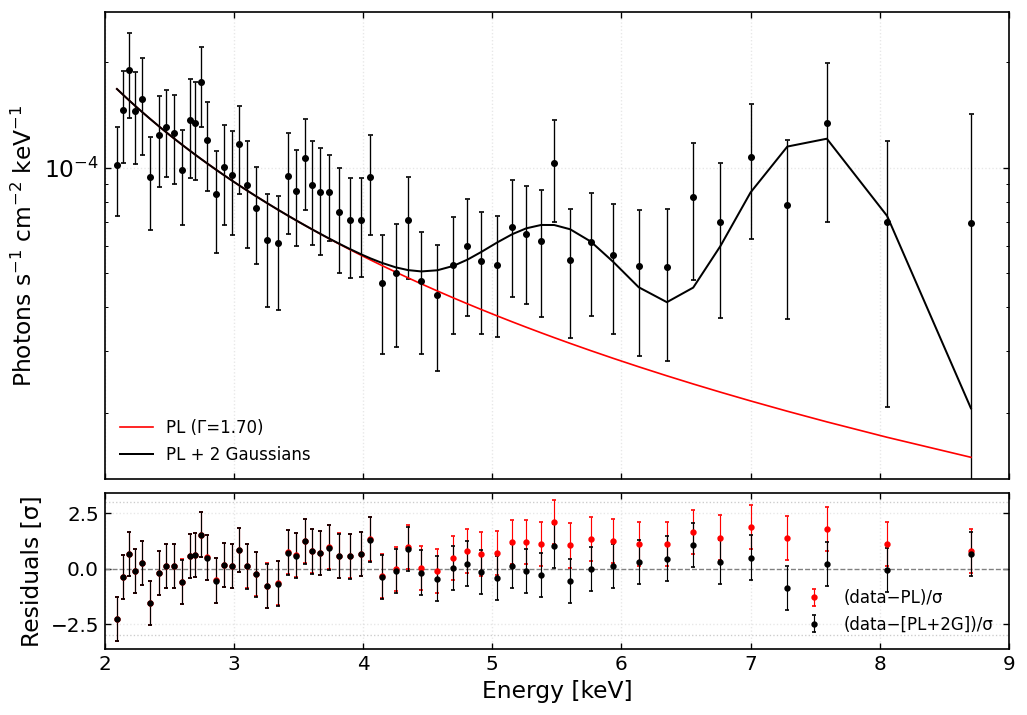}
    \caption{ESO 499-041: Chandra spectrum and data-to-model ratio in the 2.0--9.0 keV energy range fitted with a fixed power-law and two Gaussian components.}
    \label{fig:ESO499-041_Chandra}
\end{figure}
\begin{figure}
    \centering
    \includegraphics[width=0.49\linewidth]{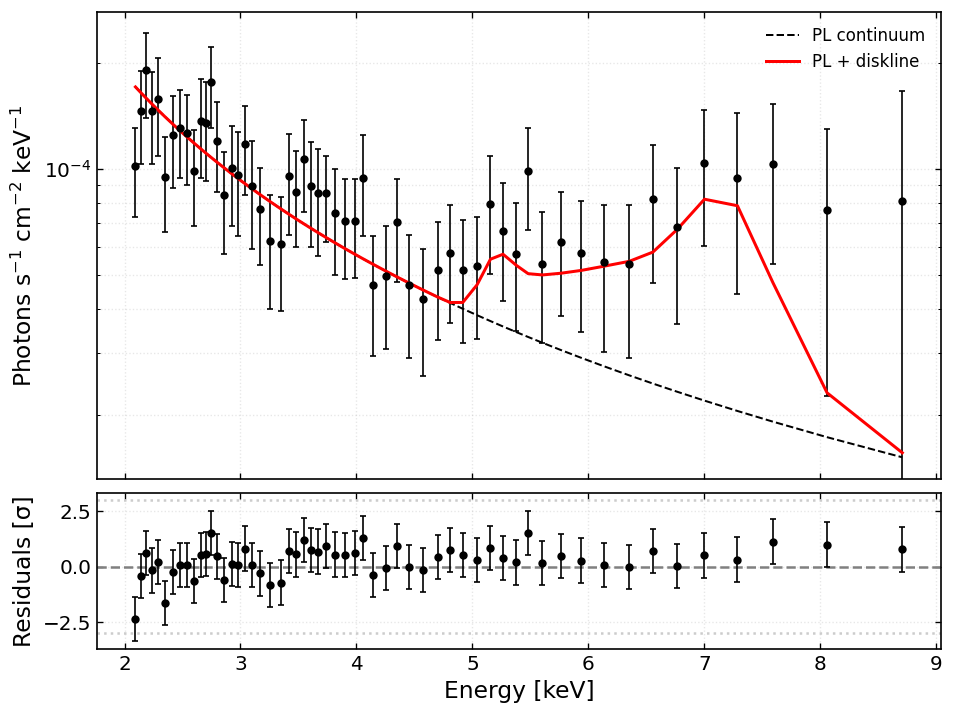}
    \includegraphics[width=0.5\linewidth]{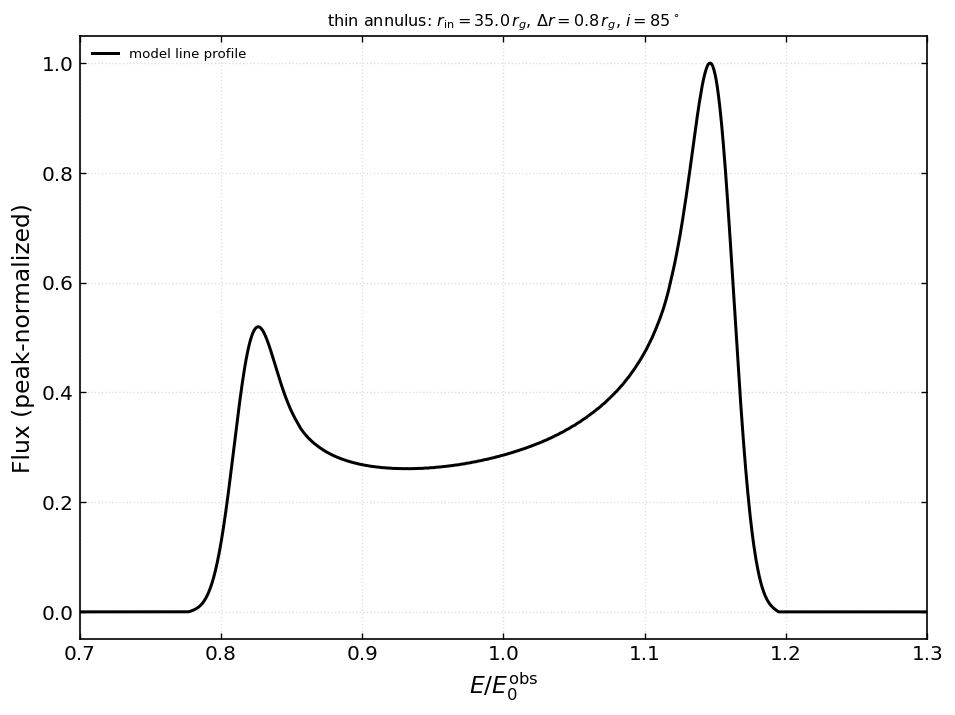} \\
    \caption{ESO 499-041: left panel --- the Chandra spectrum and data-to-model ratio in the 2.0--9.0 keV energy range fitted with the \textit{diskline} model; right panel --- the theoretical neutral iron-line profile corresponding to the adopted \textit{diskline} solution.}
    \label{fig:placeholder}
\end{figure}
While searching for archival data, we also found the Chandra observation of ESO 499-041 (ID 22249) with a 9.7 ks exposure, obtained approximately 10 years after the Swift/XRT observations (ID 00038049001 and ID 00038049002, which we used here for Model J). We obtained the source spectrum from a region with a radius of 3.6$"$, while the background spectrum was taken from an annulus with inner and outer radii of 5$"$ and 15$"$. 

The spectrum exhibits curvature beyond 5.0 keV, deviating from the power-law continuum and therefore cannot be reliably fitted with Model J. In this case, we carried out an additional inspection following the procedure outlined in the Chandra manual\footnote{https://cxc.cfa.harvard.edu/ciao/threads/flare/} to exclude the possible impact of background flare contamination on the source spectrum. The analysis revealed no evidence of flare contamination.

The spectrum in the 2–9.0 keV range was initially fitted with a power-law model with a fixed photon index of $\Gamma = 1.70$. This yielded an acceptable fit quality ($\chi^2 = 53.3$ for 56 degrees of freedom, $\chi^2_\nu = 0.95$). However, the residuals showed a clear deviation from the continuum above 5 keV. These deviations may indicate the presence of an additional component after 5 keV in the Chandra spectrum of ESO 499-041 compared to the Swift spectrum (Figure \ref{fig:ESO499-041_Chandra}). 

As the next step, we introduced two Gaussian components with fixed centroid energies corresponding to the approximate peaks in the residuals, while allowing the line widths to vary. This addition significantly improved the fit, yielding $\Delta \chi^2 = 26.08$ for $df = 4$, indicating that an additional component in the Fe-K band is statistically favoured. However, given the limited photon statistics, this result does not uniquely determine the feature's physical origin. In particular, the observed excess may correspond to a broadened relativistic profile, a blend of narrower iron lines, or other reflection-related spectral curvature. Given that ESO 499-041 is classified as a Seyfert I galaxy, a plausible explanation for this feature is the presence of neutral or ionised iron emission associated with the Fe K$\alpha$ transition, possibly shaped by Doppler boosting and gravitational redshift in the inner accretion flow. 

We explored whether this excess could be described phenomenologically by the simplest relativistic line model, accounting for the poor spectral quality above 7 keV: the \textit{diskline} model \citep{Fabian1989}, with additional constraints imposed to reduce parameter degeneracy given the limited spectral statistics. In particular, we adopted the thin-annulus approximation, thereby significantly restricting the emission region. This choice is motivated by the apparent detection of two components of the line profile — a red horn and a blue horn — with the amplitude of the blue horn exceeding that of the red one (Figure \ref{fig:placeholder}, left panel).

Assuming a cold medium and neutral iron emission, the observed energy shifts of the photons can be described using the dimensionless shift factor as $g \equiv E_{\rm obs} /E_{\rm em}$, where $E_{\rm em}$ is the rest-frame emission energy ($E_0 = 6.40$ keV for neutral Fe K$\alpha$), and $E_{\rm obs}$ is the observed photon energy. The values of $g$ encode both Doppler boosting from disc orbital motion and gravitational redshift in the deep potential well. The red and blue horns correspond to the minimum and maximum observed energy shifts, $g_{\rm red}$ and $g_{\rm blue}$, respectively. As the next step, we estimated $g_{\rm red} = 0.85$ and $g_{\rm blue} = 1.20$ from the spectrum, which correspond to observed energies: $E_{\rm red} = g_{\rm red} E_0$ and $E_{\rm blue} = g_{\rm blue} E_0$.

To provide a direct estimate of the projected orbital velocity, we can use the ratio $R \equiv g_{\rm blue} / g_{\rm red} \simeq $1.41. 
It allows us to estimate the disc inclination $i$ using parameter $\beta \sin i$, where $\beta = v/c$, as follows:
\begin{equation}
\beta \sin i = \frac{R-1}{R+1} \simeq 0.171.
\end{equation}
It is important to note that the horn factor only limits the product $\beta \sin i$, so the inclination $i$ cannot be uniquely determined from the observed line profile extrema alone. Therefore, we do not treat the inclination as a directly measured parameter, but instead explore a small grid of fixed values. Since the two peaks are clearly distinguishable (with the blue peak dominating), preference is given to the relatively large inclination. Among the possible values, angle $i=85^\circ$ provides the best overall description of the observed horn positions within the accepted assumption of a neutral line ($E_0=6.40$ keV). However, we note that $i_{\rm host}$ may differ from the inclination of the inner accretion disc. We also cannot exclude the possibility that this feature is associated with ionised iron emission, which leads to the $E_0=6.67$ keV, and this will shift the inclination to the range $i\simeq 60^\circ$--$75^\circ$.  For $i = 85^{\circ}$, this yields $\beta \simeq 0.172$. Using the Schwarzschild relation
\begin{equation}
\beta(r) = \frac{1}{\sqrt{r}\sqrt{1-2/r}},
\end{equation}
We solved for the inner emission radius $r_{\rm in}$. The outer radius was set to $r_{\rm out} = r_{\rm in} + \Delta r$, with $\Delta r = 0.8,r_g$, thereby fully defining the annular emission region. Within the restricted 2–9 keV range, this annulus model was fitted on top of the power-law continuum. All geometric parameters were fixed ($E_0$, $r_{\rm in}$, $r_{\rm out}$, $i$, $q_{\rm eff}=3$), leaving only the line normalisation $A \geq 0$ as a free parameter in addition to the power-law continuum. The best-fit annulus model yielded $\chi^2 = 31.83$ for 51 degrees of freedom ($\chi^2_\nu = 0.624$), with $r_{\rm in} = 35.0,r_g$, $i = 85^{\circ}$, continuum normalisation $K = 5.995 \times 10^{-4}$, and line normalisation $A = 9.465 \times 10^{-5}$. We note, however, that this solution should be regarded as one of the possible explanations rather than unique: it shows that a restricted relativistic-line geometry can reproduce the approximate shape of the observed Fe-K excess, but it does not constitute a formal detection of a relativistic iron line. The theoretical neutral iron line profile for the spectrum from the \textit{diskline} module is presented in Figure \ref{fig:placeholder} (right panel).

We do not regard the absence of a constrained relativistic reflection continuum as evidence against the line interpretation, rather, with the present Chandra spectrum. The continuum reflection component is not independently measurable because the data do not extend with sufficient signal-to-noise into the energy range where the reflected continuum is most diagnostic. We also note that alternative explanations remain viable. In particular, in some cases, complex absorption or partial covering can mimic an apparently broad Fe-K$\alpha$ feature (e.g., \cite{Ricci2014}), while physically motivated obscuration and reprocessing geometries may also reproduce broad Fe-K$\alpha$ structures without requiring a relativistic disc-line interpretation \citep{Yaqoob2023}. 

Given the limited photon statistics and absence of simultaneous hard X-ray coverage by Chandra and Swift observations, the present data do not permit robust discrimination among these scenarios. On the other hand, the presence of two sets of observations, separated by 10 years and showing spectral variability, prompted us to interpret the Chandra spectral feature as an Fe-K$\alpha$ line. Despite the fact that ESO 499-041 is classified as a Sy1 type galaxy, we decided to highlight this aspect. Taking into account that the Compton-thick AGNs, for which the two effects described above (complex absorption/partial covering and physically-motivated obscuration) are the most likely, albeit rare, are found among Sy1-type objects \citep{Akylas2016, Marchesi2018}. 

The case of ESO 499-041 is important for understanding the nature of nuclear activity in isolated AGNs. If the scenarios discussed in this subsection are confirmed for other isolated AGNs by forthcoming observations, this would further support the assumption that such objects do not differ fundamentally from other AGNs. In particular, hard X-ray observations with high-resolution spectroscopy, such as those anticipated for Athena, should significantly improve the diagnostics of the Fe-K features \citep{Barcons2017, Barret2023}.

\begin{table*}
\scriptsize
\setlength{\tabcolsep}{3pt}
   \centering
    \caption{Spectral models, SMBH masses, and Eddington ratio of isolated AGNs obtained from the X-ray data analysis}
    \label{tab:X-rayParam}
  \begin{tabular}{c|c|c|c|c|c|c|c|c|c|c|c|c|}
  \hline
   \hline
    2MIG & Name & $\Gamma$ &N$_{H}$& Log~L$_{2-10 keV}^{int}$ & K$_{X}$ & LogL$_{bol}$ & Spectral &$\sigma$ & $\log M_{\rm SMBH}$ &Log $L_{\rm Edd}$&Log${\lambda}_{Edd}$&Ref \\
             &      &   &   $10^{22} cm^{-2}$    &                   &         &              &     model           &km/s & [$M_\odot$] &&\\
             \hline
  1 & 2 & 3    &   4   &     5     &     6& 7           &  8 & 9& 10 &  11  & 12&   13 \\            
  \hline
    Seyfert 2 & &          &                   &         &              &              &  && &         &&    \\
  \hline
   320 & NGC 1050 & 1.99$^{+1.82}_{-1.18}$& 0.19$^{+2.19}_{-peg}$ &40.71$^{+0.21}_{-0.21}$& 15.34$^{+0.06}_{-0.06}$ & 41.93$^{+0.26}_{-0.26}$ &tbabs(zbbody + zTBabs & & 7.90$^{+0.44}_{-0.21}$ & 46.00$^{+0.44}_{-0.21}$ & -4.07$^{+0.51}_{-0.33}$& [15,19]\\ 
    & & & && & &  $\times$ cabs $\times$ zpo) & &&\\

   415 & ESO 116-018 & 1.55$^{+0.18}_{-0.15}$&313.00$^{+43.00}_{-14.00}$ & 43.30$^{+0.47}_{-0.42}$ & 16.36$^{+1.18}_{-1.09}$ & 44.51$^{+0.47}_{-0.47}$ &  MYTorus  & & 8.06$^{+0.28}_{-0.28}$ & 46.16$^{+0.28}_{-0.28}$ &-1.60$^{+0.55}_{-0.55}$&[7,15]\\ 
      & && & &&& (decoupled, edge on)  &&&\\

   417 & MCG-02-09-040 & 1.63$^{+0.11}_{-0.11}$&104.00$^{+26.00}_{-21.00}$ & 42.15$^{+0.01}_{-0.01}$ & 15.46$^{+0.06}_{-0.19}$ & 43.34$^{+0.01}_{-0.01}$ & BNtorus & 120.00$^{+9.00}_{-9.00}$& 7.07$^{+0.15}_{-0.15}$ & 45.17$^{+0.15}_{-0.15}$ &-1.83$^{+0.15}_{-0.15}$&[1,10]\\
      & & & & & & && &&\\

   1086 &  IC 2227 & 1.86$^{+0.20}_{-0.14}$ &60.00$^{+11.00}_{-7.00}$  &41.89$^{+0.09}_{-0.38}$&   15.71$^{+0.23}_{-0.44}$ &  43.92$^{+0.24}_{-0.24}$& MYTorus    & 155.50$^{+3.60}_{-3.60}$ &7.70$^{+0.05}_{-0.05}$ & 45.80$^{+0.05}_{-0.05}$ &-2.73$^{+0.25}_{-0.25}$&[8,9]\\ 
    & & & & & & & &&&\\ 

    1363 & NGC 3081 & 1.75$^{+0.03}_{-0.03}$ &66.00$^{+2.00}_{-2.00}$ &42.78$^{+0.01}_{-0.01}$ & 15.75$^{+0.05}_{-0.48}$ & 43.98$^{+0.01}_{-0.01}$ &  MYTorus  &  $129^{+8.00}_{-8.00}$ & 7.25$^{+0.13}_{-0.13}$ & 45.35$^{+0.13}_{-0.13}$ & -1.37$^{+0.13}_{-0.13}$& [2,13]\\
      & & && & & &(decoupled, face on) &&\\

    1442 & ESO 317-038 & 1.56$^{+0.14}_{-0.15}$ &86.60$^{+27.60}_{-22.60}$&41.98$^{+0.10}_{-0.21}$ & 15.43$^{+0.07}_{-0.16}$ & 43.17$^{+0.01}_{-0.01}$ & $\rm{ C_{1}({C_{2} \times zpo} + zTBabs } $ &152.00$^{+3.00}_{-3.00}$& 7.65$^{+0.05}_{-0.05}$& 45.75$^{+0.05}_{-0.05}$ &-2.58$^{+0.05}_{-0.05}$&[10,19] \\ 
     & & & & && & $\rm{\times zpo +{C_{3} \times pexmon})} $ &&&&& \\

   1607 & IGR J11366-6002&1.79$^{+0.04}_{-0.04}$  &  1.37$^{+0.52}_{-0.39}$ &42.14$^{+0.01}_{-0.01}$ &     15.46$^{+0.08}_{-0.18}$   & 43.33$^{+0.10}_{-0.10}$ &  xillver $+$ cutoffpl  & & 6.59$^{+0.10}_{-0.10}$ & 44.69$^{+0.10}_{-0.10}$ &-1.41$^{+0.14}_{-0.14}$&[1,17]\\ 
      & & & & & & & &&&&&\\

   1915 & NGC 5347 & 1.60$^{+0.37}_{-0.27}$ & &41.46$^{+0.03}_{-0.03}$ & 15.36$^{+0.06}_{-0.09}$ & 42.65$^{+0.06}_{-0.09}$ & zbbody + pexmon &  $89.80^{+2.86}_{-2.86}$ & 6.36$^{+0.08}_{-0.08}$ & 44.46$^{+0.08}_{-0.08}$ &-1.81$^{+0.10}_{-0.12}$&[1,11] \\ 
    & & & & & & & &&&&&\\

    1950 & ESO 097-013 & 1.68$^{+0.05}_{-0.09}$ & 595.00$^{+76.00}_{-60.00}$&42.80$^{+0.03}_{-0.04}$ & 15.77$^{+0.06}_{-0.05}$ & 43.99$^{+0.03}_{-0.03}$ &  XCLumpy &  $102.00^{+7.00}_{-7.00}$ & 6.67$^{+0.14}_{-0.14}$ & 44.77$^{+0.14}_{-0.14}$ &-0.77$^{+0.14}_{-0.14}$&[3,10] \\ 
      & & & & & & & &&&\\

    2363 & NGC 6300 & 1.84$^{+0.04}_{-0.03}$ & 21.80$^{+0.40}_{-0.03}$&42.14$^{+0.02}_{-0.02}$ & 15.46$^{+0.06}_{-0.19}$ & 43.33$^{+0.02}_{-0.02}$ &  MYTorus  &  $81.00^{+2.00}_{-2.00}$ & 6.11$^{+0.09}_{-0.09}$ & 44.21$^{+0.09}_{-0.09}$ &-0.88$^{+0.09}_{-0.09}$& [5,10] \\
      & & & && && (decoupled, face on)&& \\

    2811 & NGC 6951 & 2.13$^{+0.83}_{-0.88}$ & \textbf{4.0$^{+2.7}_{-1.9}$} & 40.04$^{+0.14}_{-0.14}$ & 15.33$^{+0.06}_{-0.06}$ & 41.23$^{+0.14}_{-0.14}$  & zbbody + zTBabs $\times$ zpo  &  $115.00^{+4.00}_{-4.00}$  & 6.96$^{+0.08}_{-0.08}$ & 45.06$^{+0.08}_{-0.08}$ &-3.83$^{+0.16}_{-0.16}$& [14,19] \\ 
      & & & & & & & &&&&&\\

    3118 & NGC 7479 & 1.83$^{+0.09}_{-0.09}$ &363.60$^{+58.00}_{-50.30}$ &42.64$^{+0.14}_{-0.09}$ & 15.66$^{+0.11}_{-0.39}$ & 43.83$^{+0.11}_{-0.11}$  & MYTorus  &  $144.52^{+3.51}_{-3.51}$ & 7.52$^{+0.05}_{-0.05}$ & 45.62$^{+0.05}_{-0.05}$ &-1.79$^{+0.12}_{-0.12}$&[7,14] \\
      & & & && & &(decoupled, edge on)  &&&&\\

   \hline
    Seyfert 1 &      &          &                   &         &              &                &  &&&  &&\\
    \hline

    1550 & ESO 438-009 & 1.78$^{+0.05}_{-0.08}$ & &42.67$^{+0.01}_{-0.01}$ & 15.68$^{+0.05}_{-0.41}$ & 43.86$^{+0.01}_{-0.01}$ &  pexmon + cutoffpl &  & 7.98$^{+0.45}_{-0.45}$ & 46.08$^{+0.45}_{-0.45}$ &-2.22$^{+0.45}_{-0.45}$&[1,15] \\
      & & & & & & & &&&&&\\ 

    1646  &  CGCG 243-024& 1.76$^{+0.07}_{-0.02}$& &42.23$^{+0.03}_{ -0.03}$ & 15.48$^{+0.06}_{-0.21}$   & 43.42$^{+0.03}_{-0.03}$  & $\rm{zbbody+mekal}$&& 6.40$^{+0.11}_{-0.11}$ & 44.50$^{+0.11}_{-0.11}$ &-1.08$^{+0.11}_{-0.11}$&[18,19]  \\
    & & & & & & & $\rm{+nthComp}$&&&&&\\ 

    1873 & NGC 5231 & 1.67$^{+0.05}_{-0.04}$ & 2.30$^{+0.13}_{-0.14}$&42.91$^{+0.02}_{-0.02}$ & 15.86$^{+0.72}_{-0.59}$ & 44.11$^{+0.5}_{-0.5}$ & $\rm{C_{1}(apec + zTBabs \times}$ & $143.55^{+2.49}_{-2.49}$ & 7.51$^{+0.04}_{-0.04}$ & 45.61$^{+0.04}_{-0.04}$ &-1.50$^{+0.50}_{-0.50}$&[9,19] \\ 
      & & && & & &$ (\rm{zpo} + \rm{C_{2} \times xillver))}$ &&&&& \\

    2183 & UGC 10120 & 2.14$^{+0.03}_{-0.03}$& & 42.26 & 15.49$^{+0.05}_{-0.22}$ & 43.45 &   tbabs $\times(\rm{po+}$ & & 6.14$^{+0.04}_{-0.04}$ & 44.24$^{+0.04}_{-0.04}$ &-0.79$^{+0.04}_{-0.04}$&[4,16] \\ 
    & & && & & &$\rm{kdblur2} \times \rm{reflionx}$) & &&&&\\

    3128 & IC 5287 & 1.53$^{+0.22}_{-0.22}$ & 0.39$^{+0.12}_{-0.11}$&42.09$^{+0.10}_{-0.10}$&     15.45$^{+0.08}_{-0.18}$    & 43.28$^{+0.10}_{-0.10}$   &   $\rm{zTBabs\times zpo}$ & $142.10^{+3.90}_{-3.90}$ & 7.48$^{+0.06}_{-0.06}$ & 45.58$^{+0.06}_{-0.06}$ &-2.30$^{+0.12}_{-0.12}$&[9,19]      \\ 
           & & & & & & &&&& &&\\

     384 & ESO 499-041 &1.78$^{+0.29}_{-0.22}$&&$42.57^{+0.05}_{-0.05}$&15.62$^{+0.07}_{-0.35}$    & 43.76$^{+0.05}_{-0.05}$&$ \rm{zbremss}+$&  & 8.28$^{+0.44}_{-0.44}$ & 46.38$^{+0.44}_{-0.44}$ &-2.62$^{+0.44}_{-0.44}$&  [15,19]\\
      & & & & & & &$+\rm{zTBabs} \times \rm{zpo}$&&&&&\\

    1126 & CGCG 179-005 & 2.02$^{+0.04}_{-0.04}$& 0.38$^{+0.22}_{-0.22}$& $42.34^{+0.09}_{-0.02}$ & 15.50$^{+0.06}_{-0.24}$ & 43.50$^{+0.02}_{-0.02}$ &  $\rm{TBpcf \times zpo +}$ &  $98.70^{+12.0}_{-12.0}$ &6.59$^{+0.21}_{-0.21}$ & 44.69$^{+0.21}_{-0.21}$ &-1.60$^{+0.21}_{-0.21}$&[12,19] \\ 
     & & & & & & & $+ zgauss$&&&&&\\

    1516 & ESO 215-014 & 1.58$^{+0.14}_{-0.13}$ & &42.69$^{+0.06}_{-0.06}$ &      15.69$^{+0.08}_{-0.42}$ & 43.89$^{+0.05}_{-0.05}$  & $\rm{zpo}$    &  & 7.94$^{+0.44}_{-0.44}$ & 46.04$^{+0.44}_{-0.44}$ &-2.15$^{+0.44}_{-0.44}$&[15,19]  \\  
      & & & & & & & &&&&&\\
           \hline
            \end{tabular}
\begin{minipage}{\linewidth}
\smallskip
NOTE: 
Column~1: Galaxy number in the 2MIG catalogue;
Column~2: Galaxy`s name;
Column~3: Photon index $\Gamma$.
Column~4: Column density N$_{H}$, cm$^{-3}$
Column~5: Intrinsic luminosity $\log L_{2-10\,{\rm keV}}^{\rm int}$;
Column~6: Bolometric correction $K_X$;
Column~7: $\log{L_{bol}}$;
Column~8: Adopted spectral model;
Column~9: Stellar central velocity dispersion $\sigma_\star$ (km\,s$^{-1}$);
Column~10: $\log{M_{BH}}$ derived from $\sigma_\star$;
Column~11: $\log{L_{\rm Edd}}$;
Column~12: $\log\lambda_{\rm Edd}$;
Column~13: reference.
The Photon indexes, intrinsic luminosity L$_{2-10 keV}^{int}$ and spectral models were taken from the articles as follows by Refs: (1) -- \cite{Vasylenko2020} Table 2: NGC 5347 -- pexmon model, MCG-02-09-040 -- BNTorus model, J11366-6002 -- xillver model; (2) -- \cite{Traina2021}, Table 3: MYTorus model, de-coupled regime, face-on; (3) -- \cite{Uematsu2021}, Table 4, baseline model -- XCLumpy; (4) -- \cite{2018MNRASBonson}, used parameters of the XMM15a model \rem{luminosity for XMM15a model (high level of variability for photon index and luminosity + relativistic iron line)}; (5) -- \cite{Jana2020}, Table 6: S1, (variability); (6) -- \cite{2009A&AAkylas}; (7) -- \cite{2019Marchesi} Table 4: ESO 116-018, Table 2: NGC 7479, Table 4: MYTorus, de-coupled, face-on; (8) -- \cite{Silver_2022}, Table 5; the velocity dispersions $\sigma$ are taken from the SDSS -- (9) \cite{2017SDSS} and individual articles. The central velocity dispersions for MCG-02-09-040, ESO 317-038, ESO 097-013, and NGC 6300 -- (10) \cite{2022ApJS2616K}; NGC 5347 -- (11) \cite{1990MNRASTerlevich}; CGCG 179-005 and MCG+09-25-022 -- (12) \cite{2006ApJGreene}; NGC 3081 -- (13) \cite{2005MNRASGarcia-Rissmann};  NGC 6951 and NGC 7479 -- (14) -- \cite{2009ApJHo}. The SMBH masses of ESO 116-018, ESO 438-009, ESO 215-014, ESO 499-041, and NGC 1050 were taken from the catalogue compiled as an extension of the NANOGrav Programme for Gravitational Waves and Fundamental Physics. -- (15) \cite{2019BAAS51g}, and references therein search for potential double SMBHs \citep{2021ApJ914121A}. The catalogue compiled in the latter article contains SMBH masses obtained by several methods. We utilised the mass estimates obtained by a single method, which yielded the largest increase in the number of objects for analysis. Namely, we exploited the M$_{SMBH}$-M$_{bulge}$ relation using the correlation by \cite{2013ApJ764184M}). The SMBH masses of UGC 10120 are obtained by -- (16) \cite{2019ApJ87723L}, IGR J11366-6002 -- (17) \cite{2022ApJS2612K}, CGCG 243-024 -- (18), \cite{2018MNRAS4771086H}, spectral model used in this article -- (19).
        \end{minipage}            
                \end{table*}
\section{SMBH masses and Eddington ratio of isolated AGNs}
 \label{sec:SMBHmass}
 The SMBH mass, accretion rate, and Eddington ratio are the basic parameters of AGNs. In the case of isolated AGNs, their host galaxies evolve under the limited influence of the neighbouring environment, in contrast to galaxies in groups/clusters, where the environment is conducive to mergers and to maintaining nuclear activity. Should the AGN parameters differ between isolated galaxies and galaxies in a tight environment? Exploring this aspect, we determined the SMBH masses of isolated AGNs, setting an approximate upper limit to AGN energetics via the Eddington limit.

\subsection{SMBH masses}

For host galaxies with measured stellar central velocity dispersion $\sigma_\star$, we estimated the SMBH masses using the $M_{\rm BH}$--$\sigma_\star$ relation of \citet{2013ApJMcConnell}:
\begin{equation}
\log \left(\frac{M_{\rm BH}}{M_\odot}\right) = \alpha + \beta \log\left(\frac{\sigma_\star}{200~{\rm km~s^{-1}}}\right),
\end{equation}
with $\alpha=8.32$ and $\beta=5.64$ (consistent with Eq.~(7) in the \citep{2013ApJMcConnell}).
The uncertainties for $\log M_{\rm BH}$ were calculated by propagating the measurement uncertainty for $\sigma_\star$ and the quoted uncertainties of $(\alpha,\beta)$. Also, we added the intrinsic scatter $\epsilon_0$ of the adopted $M_{\rm BH}$--$\sigma_\star$ relation in quadrature:
\begin{equation}
\begin{aligned}
\Delta\log_{10} M_{\rm BH} =
\Big[&(\Delta a)^2 + \big(\Delta b\,\log_{10}(\sigma/200)\big)^2 \\
&+ \left(\frac{b}{\ln 10}\,\frac{\Delta\sigma}{\sigma}\right)^2
+ \mathrm{scatter}^2 \Big]^{1/2}.
\end{aligned}
\end{equation}
For a small subset of isolated AGNs in our sample that lack reliable $\sigma_\star$ measurements, we used published SMBH mass estimates and their uncertainties when available (see the notes to Tables~2 and~3). 

Totally, the 32 isolated AGNs have the following SMBH mass distribution: 9/32 objects in $10^6 \le M_{\rm BH}/M_\odot < 10^7$, 18/32 in $10^7 \le M_{\rm BH}/M_\odot < 10^8$, 4/32 in $10^8 \le M_{\rm BH}/M_\odot < 10^9$, and 1/32, NGC 7749, in $10^9 \le M_{\rm BH}/M_\odot < 10^{10}$.

\subsection{Eddington ratio}

The sample of 2MIG isolated AGNs includes galaxies with different activity types. To obtain consistent bolometric luminosities across the sample, we adopted the hard X-ray bolometric correction by \citet{2020A&ADuras}, derived from the analysis of $\sim$1000 Seyfert~1 and Seyfert~2 AGNs. \citet{2020A&ADuras} provided both a universal correction applicable to the full AGN sample and type-dependent corrections; these prescriptions are consistent within the overlapping luminosity range. Their results show that the hard X-ray bolometric correction is relatively constant at low luminosities and increases with luminosity. We computed the X-ray bolometric correction following \cite{2020A&ADuras}:
\begin{equation}
K_{X} = a \left[1+ \frac{\log (L_{2-10\,{\rm keV}}^{\rm int}/L_{\odot})}{b}\right]^{c},
\end{equation}
where the coefficients $a$, $b$, and $c$ are taken from Table~1 by \cite{2020A&ADuras}. To derive $L_{\rm bol}$ and the Eddington ratio, we compiled intrinsic (absorption-corrected) 2--10~keV luminosities, $L_{2-10\,{\rm keV}}^{\rm int}$, obtained from the best-fit X-ray spectral models (this work and/or literature). The resulting parameters are listed in Table~\ref{tab:X-rayParam}, which summarises the X-ray spectral properties of isolated AGNs and allows us to test possible trends between the photon index $\Gamma$, $L_{2-10\,{\rm keV}}^{\rm int}$, and the Eddington ratio $\lambda_{\rm Edd}$.

Thus, the Eddington ratio is defined as follows:
\begin{equation}
\lambda_{\rm Edd} \equiv \frac{L_{\rm bol}}{L_{\rm Edd}}
= \frac{K_{X}\,L_{2-10\,{\rm keV}}^{\rm int}}{L_{\rm Edd}},
\end{equation}
where $L_{2-10\,{\rm keV}}^{\rm int}$ is the intrinsic (absorption-corrected) 2--10~keV luminosity, and
\begin{equation}
L_{\rm Edd}=1.26\times10^{38}\left(\frac{M_{\rm BH}}{M_\odot}\right)\,{\rm erg\,s^{-1}}.
\end{equation}
Table~\ref{tab:general} contains observed fluxes $F_{\rm obs}$ for reference; these observed values are not used directly to compute $\lambda_{\rm Edd}$.
\begin{table}
\scriptsize
\setlength{\tabcolsep}{4pt}
\caption{The SMBH masses for isolated AGNs, which were obtained without detected X-ray emission}
\label{tab:SMBHmass}
\begin{tabular}{c|c|c|c|c|c}
\hline
\hline
2MIG & Name & $\sigma$ & $\log M_{SMBH}$ & Log~$L_{Edd}$ & Ref \\
     &      & km/s & $[\log_{10}(M_\odot)]$ &  & \\
\hline
Seyfert 2 & & & & & \\
\hline
267 & UGC 01757 & 143.00$^{+20.00}_{-20.00}$ & 7.50$^{+0.52}_{-0.52}$ & 45.60$^{+0.52}_{-0.52}$ & [2]\\
& & & & & \\
1454 & MCG-02-27-009 & 142.20$^{+5.90}_{-5.90}$ & 7.48$^{+0.40}_{-0.40}$ & 45.58$^{+0.40}_{-0.40}$ & [4]\\
& & & & & \\
1571 & UGC 06398 & 204.63$^{+5.25}_{-5.25}$ & 8.38$^{+0.39}_{-0.39}$ & 46.48$^{+0.39}_{-0.39}$ & [1]\\
& & & & & \\
\hline
Seyfert 1 & & & & & \\
\hline
1633 & UGC 06769 & 158.81$^{+4.90}_{-4.90}$ & 7.76$^{+0.39}_{-0.39}$ & 45.86$^{+0.39}_{-0.39}$ & [1]\\
& & & & & \\
2067 & MCG+09-25-022 & 139.00$^{+17.00}_{-17.00}$ & 7.43$^{+0.49}_{-0.49}$ & 45.53$^{+0.49}_{-0.49}$ & [3]\\
\hline
LINER & & & & & \\
\hline
1914 & MCG-03-35-020 & 164.00$^{+14.60}_{-14.60}$ & 7.83$^{+0.44}_{-0.44}$ & 45.93$^{+0.44}_{-0.44}$ & [5]\\
& & & & & \\
1989 & PGC 989455 & 206.10$^{+31.60}_{-31.60}$ & 8.39$^{+0.54}_{-0.54}$ & 46.49$^{+0.54}_{-0.54}$ & [5]\\
& & & & & \\
2357 & UGC 10774 & 96.70$^{+6.33}_{-6.33}$ & 6.54$^{+0.43}_{-0.43}$ & 44.64$^{+0.43}_{-0.43}$ & [1]\\
& & & & & \\
2509 & PGC 206329 & 172.90$^{+23.80}_{-23.80}$ & 7.96$^{+0.51}_{-0.51}$ & 46.06$^{+0.51}_{-0.51}$ & [5]\\
& & & & & \\
\hline
AGN & & & & & \\
\hline
2018 & CGCG 248-019 & 137.35$^{+4.75}_{-4.75}$ & 7.40$^{+0.40}_{-0.40}$ & 45.50$^{+0.40}_{-0.40}$ & [1]\\
& & & & & \\
2202 & UGC 10244 & 133.48$^{+5.25}_{-5.25}$ & 7.33$^{+0.40}_{-0.40}$ & 45.43$^{+0.40}_{-0.40}$ & [1]\\
& & & & & \\
3128 & NGC 7749 & 285.50$^{+29.00}_{-29.00}$ & 9.19$^{+0.46}_{-0.46}$ & 47.29$^{+0.46}_{-0.46}$ & [5]\\
& & & & & \\
\hline
\hline
\end{tabular}
\begin{minipage}{\linewidth}
\smallskip
NOTE: Column 1: number of a galaxy in the 2MIG catalogue. Column 2: Galaxy's name. Column 3: central velocity dispersion, in km/s. Column 4: SMBH mass in logarithmic units, $\log_{10}(M_{\rm SMBH}/M_\odot)$, derived using the $M$--$\sigma$ relation with uncertainty propagation and intrinsic scatter. Column 5: Eddington luminosity computed as $\log L_{\rm Edd} = 38.10 + \log M_{\rm SMBH}$ (erg s$^{-1}$); thus $d\log L_{\rm Edd} = d\log M_{\rm SMBH}$. Column 6: reference.
The velocity dispersions $\sigma$ are taken from the SDSS -- (1) \cite{2017SDSS}. The SMBH masses are estimated for UGC 01757 -- (2) \cite{1990MNRASTerlevich}; MCG+09-25-022 -- (3) \cite{2006ApJGreene}; MCG-02-27-009 -- (4) \cite{2003AJWegner}; MCG-03-35-020, PGC 989455, PGC 206329, and NGC 7749 -- (5) \cite{2014MNRASCampbell}.
\end{minipage}
\end{table}

\section{Main relationships: Discussion}
 \label{sec:discussion}

\subsection{General X-ray properties of isolated AGNs}
\begin{figure}
    \centering
    \includegraphics[width=0.85\linewidth]{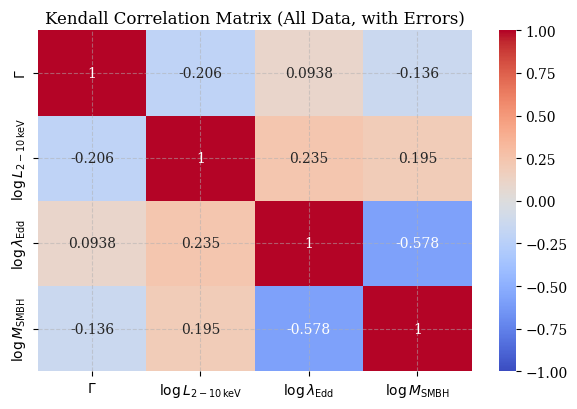}
  \caption{The Kendall correlation matrix for relationships between X-ray properties of the 2MIG isolated AGNs}
   \label{fig:cross}
\end{figure}
\begin{figure}
    \centering
    \includegraphics[width=0.80\linewidth]{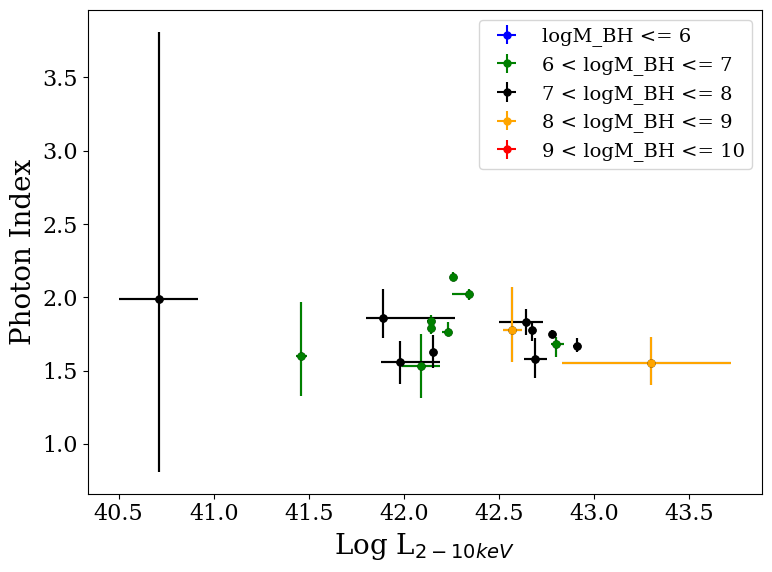}
    \caption{Relation between the Photon Index and X-ray luminosity in 2-10 keV for 2MIG isolated AGNs.}
    \label{fig:PhoI_ratio}
\end{figure}
Our sample includes 61 isolated galaxies, where only 25 were observed in the X-ray range (Table \ref{tab:general}). Moreover, of these 25 galaxies, only 19 have high-quality spectra (Table \ref{tab:X-rayParam}. This limited number of detections precludes comprehensive statistical analysis; however, it provides a foundation for examining the general properties of isolated AGNs. 

Due to the small sample size and the fact that the data do not follow a normal distribution, we first applied the Kendall's rank correlation coefficient. This non-parametric method was used to assess whether there were statistically significant trends of monochromatic growth or decline in the selected parameters. 

The correlation significance was assessed using an uncertainty-aware Monte Carlo resampling approach to obtain corresponding $p$-values, which quantify the statistical significance of the observed trends. Following the general idea of resampling methods \citep{Shaqiri2023}, we generated repeated realisations of the original dataset by perturbing each measurement according to its quoted $1\sigma$ uncertainty.
$x_i' = x_i + \mathcal{N}(0,\sigma_{x,i})$ and $y_i' = y_i + \mathcal{N}(0,\sigma_{y,i})$, and recomputed the Kendall rank coefficient for each realisation. This procedure yields an empirical distribution of the correlation coefficient and allows us to estimate confidence intervals (95\%) and $p$-values. For our small sample of isolated AGNs, propagating heteroscedastic measurement uncertainties in this way improves the robustness of the correlation analysis. It reduces the impact of potential bias driven by a few highly uncertain sources.

We constructed a correlation matrix (Fig.~\ref{fig:cross}) of the general X-ray properties of the isolated AGNs sample, propagating the measurement uncertainties via Monte Carlo resampling. The Kendall coefficients (with the corresponding $p$-values) are: $\tau=-0.136$ with $p=0.57$ between the photon index $\Gamma$ and $\log M_{\mathrm{SMBH}}$; $\tau=-0.206$ with $p=0.228$ between $\Gamma$ and $\log L_{2-10\,\mathrm{keV}}$; $\tau=0.195$ with $p=0.308$ between $\log L_{2-10\,\mathrm{keV}}$ and $\log M_{\mathrm{SMBH}}$; and $\tau=0.0938$ with $p=0.622$ between $\Gamma$ and $\log\lambda_{\mathrm{Edd}}$. These $p$-values indicate that none of the above relations is statistically significant at typical thresholds (e.g., $p<0.05$) for the full sample; therefore, no robust correlation can be claimed based on the global dataset alone.

To further test whether any apparent trends are driven by a small number of highly uncertain points and to jointly account for both horizontal and vertical uncertainties, we additionally performed Bayesian linear regression using \texttt{linmix}. For all relations involving $\Gamma$ (i.e., $\Gamma$ versus $\log L_{2-10\,\mathrm{keV}}$, $\log\lambda_{\mathrm{Edd}}$, $\log M_{\mathrm{SMBH}}$, and $\log(L_{2-10}/L_{\mathrm{Edd}})$), the posterior slope distributions are consistent with zero within the 95\% credible intervals. Likewise, for $\log L_{2-10\,\mathrm{keV}}$ versus $\log M_{\mathrm{SMBH}}$ the inferred slope is not significantly different from zero for the full sample ($\beta=0.229$, 95\% CI $[-0.409,\,0.871]$), with a substantial intrinsic scatter ($\sigma_{\rm int}\simeq 0.62$). This confirms that, once measurement uncertainties and intrinsic scatter are accounted for, the global sample provides no statistically robust evidence for monotonic correlations among these parameters.

The correlation between $\Gamma$ and accretion state indicators in such a wide range of luminosities/Eddington ratios cannot usually be described by a single monotonic dependence consistent with previous results. Instead, the previous work shows that the sign and slope of the correlation $\Gamma$--$l_X$ (where $l_X\equiv L_X/L_{\rm Edd}$) or $\Gamma$--$\lambda_{\rm Edd}$ (where $\lambda_ {\rm Edd}\equiv L_{\rm bol}/L_{\rm Edd}$) depend on the accretion regime: for relatively high values of $\lambda_{\rm Edd}$ (or $l_X$), a positive correlation is more often observed, while in low-luminosity AGN with small $\lambda_ {\rm Edd}$ values, an anticorrelation $\Gamma$ with $\lambda_{\rm Edd}$ / $l_X$ is found \cite[e.g.][]{Shemmer2006, Shemmer2008, Risaliti2009, Brightman2013, GuCao2009, Younes2011}.

\subsection{Relation between the Photon index and Log L$_{2-10keV}$/L$_{Edd}$}

\begin{figure}
    \centering
    \includegraphics[width=0.9\linewidth]{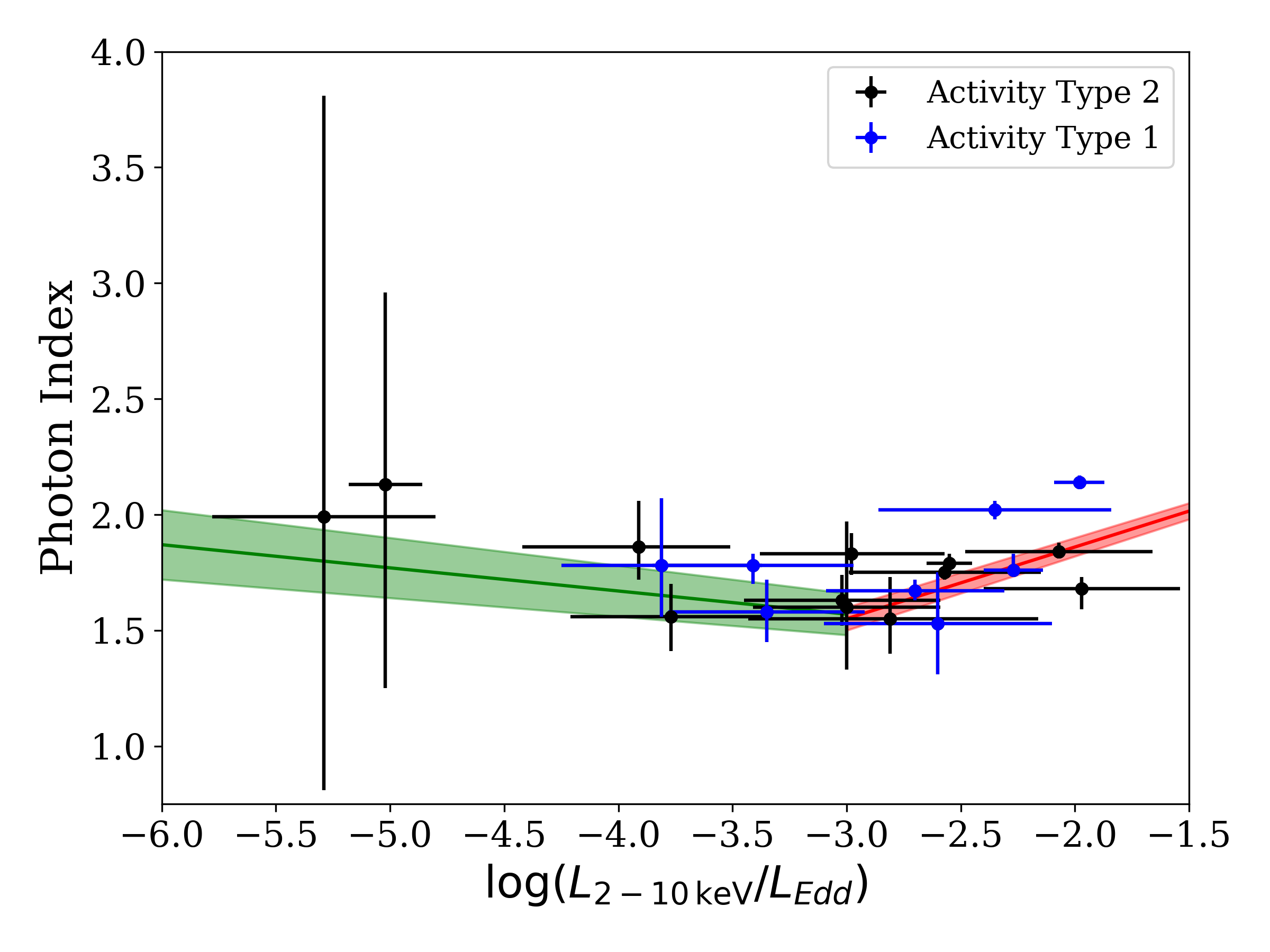}
    \caption{Relation between the Photon index and Log L$_{2-10keV}$/Ledd for 2MIG isolated AGNs, where bolometric luminosity was calculated with $K_{X}$ corrections.} 
    \label{fig:PhoI_accr}
\end{figure}

To place our isolated AGNs into a physically motivated context, we compare their locations in the $\Gamma$--$l_X$ plane with the regime-dependent picture established for much larger and more heterogeneous AGN/XRB samples. We stress that this comparison is used primarily as a qualitative diagnostic of accretion regimes, rather than as an attempt to re-derive the full piecewise $\Gamma$--$l_X$ correlations from our limited 2MIG dataset. In particular, \cite{2015MNRASYang} exploited a large sample of AGNs with different types of activity and $l_{X}=L_{2-10 keV}$/$L_{Edd}$ in the range of $10^{-9}$ to $10^{-1}$. It allowed them to conclude that $\Gamma$ is positively and negatively correlated with $l_{X}$ when $l_{X} \geq 10^{-3}$ and $10^{-6.5} \leq lX \leq 10^{-3}$, respectively, while $\Gamma$ is almost constant when $l_{X} \leq 10^{-6.5}$. These authors explained the above correlation in the framework of a coupled hot-accretion flow-jet model. Radio emission always comes from the jet, while X-ray emission comes from the accretion flow and jet when $l_{X}$ is above and below $10^{-6.5}$, respectively. More specifically, they assumed that as the accretion rate increases, the hot accretion flow develops into a clumpy, disc-corona two-phase structure due to thermal instability. They also assumed that such a kind of two-phase accretion flow could explain the observed positive correlation, which is for $l_{X} > 10^{-3}$:
\begin{equation}
    \Gamma =  (0.31 \pm 0.01)~log_{10}(L_{X}/L_{Edd}) + (2.48 \pm 0.02)
\end{equation}
and a negative correlation for the moderate-luminosity branch $10^{-6.5} \leq l_{X} \leq 10^{-3}$:
 \begin{equation}
    \Gamma = (0.1 \pm 0.02)~log_{10}(L_{X}/L_{Edd}) + (1.27 \pm 0.03).
\end{equation}
We have plotted this relation between the Photon index and Log L$_{2-10keV}$/Ledd for isolated AGNs in Fig. \ref{fig:PhoI_accr}. Given the modest sample size and the restricted $l_X$ coverage of 2MIG, we do not expect to statistically recover the full piecewise $\Gamma$-$l_X$ trends. Instead, we use the $\Gamma$--$l_X$ diagram to interpret individual objects in terms of plausible accretion regimes, while our formal correlation analysis (Monte--Carlo Kendall and \texttt{linmix}) indicates that any global $\Gamma$--$l_X$ trend within the 2MIG sample is not statistically significant. One can see that eight galaxies (NGC 6951, NGC 1050, IC 5287, IC 2227, ESO 499-041, ESO 317-038, ESO 215-014, and NGC 5347) have fallen in the green region (Fig. \ref{fig:PhoI_accr}).

According to \cite{2015MNRASYang}, it can be explained as an accretion regime of ''ADAF + type I LHAF" (Advection-Dominated Accretion Flow, Luminous Hot Accretion Flow), when the X-ray emission comes from a one-phase hot-accretion flow. The one-phase hot model describes the region where synchrotron photons (soft X-rays) undergo self-absorption. That is, here the soft radiation is artificially reduced while the hard radiation is maintained; the spectral index can be approximately 1.5 or 1.6.

The X-ray parameters of NGC 5231, NGC 3081, ESO 116-018, CGCG 243-024, and NGC 6300 have fallen into the red region (Fig. \ref{fig:PhoI_accr}), which means that these AGNs have a two-phase accretion flow. In other words, the Compton up-scattering of the optical/ultraviolet photons from the cold clumps will effectively cool the electrons in the hot phase. 

Such objects as ESO 438-009, NGC 7479, IGRJ 11366-6002, CGCG 179-005, UGC 10120, and ESO 097-013, do not exactly belong to any specific region. Except for CGCG 179-005 and UGC 10120, all are classified as Sy2 type and their spectra are well described by the models with a clumpy torus component (see Table \ref{tab:X-rayParam}). We recall that clumpiness in the dusty torus structure complicates the evaluation of intrinsic luminosity, leading to observable deviations in its properties and making the determination of luminosity ratios more difficult. CGCG 179-005, classified as a BLAGN, exhibits an X-ray spectrum modelled as a redshifted power law with a partial covering factor. UGC 10120, also known as Mrk 493, turned out to have relativistic effects in its spectrum for the accretion disc reflection component, which are very difficult to take into account when understanding the primary power law index \citep{2018MNRASBonson}.

\subsection{Column density in isolated AGNs }

Another important characteristic of AGNs is the line-of-sight X-ray absorbing column density, $N_{\rm H}$. \citet{Ricci2017} showed that the cumulative $N_{\rm H}$ distribution depends on the merger stage of the host galaxy, with late-stage mergers hosting systematically more heavily obscured AGN.

\begin{figure}
    \centering
    \includegraphics[width=1.0\linewidth]{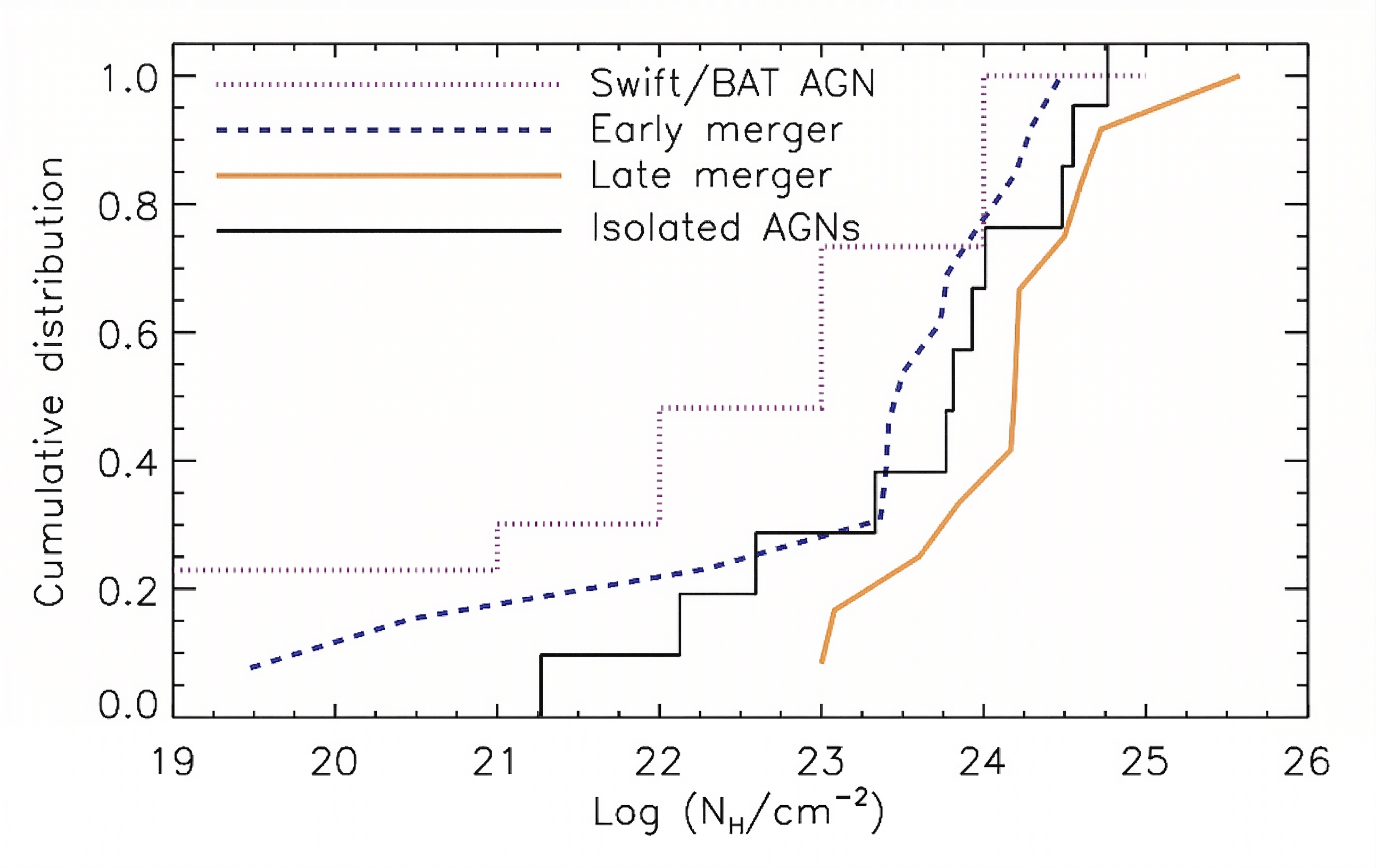}
    \caption{Cumulative line-of-sight ($N_{\rm H}$) distribution. The purple dotted, blue dashed, and orange solid curves show the Swift/BAT AGN, early-merger, and late-merger samples from \citet{Ricci2017}, respectively. The solid black curve shows the cumulative $N_{\rm H}$ distribution for the studied 2MIG isolated AGNs (11 Seyfert~2 galaxies with available $N_{\rm H}$ measurements).}
\label{fig:Nh}
\end{figure}

The $N_{\rm H}$ measurements are available for only 11 Seyfert~2 galaxies from our sample. For these objects, we constructed the cumulative \(N_{\rm H}\) distribution following the same approach as \citet{Ricci2017} and overplotted it on their published comparison plot (see Fig \ref{fig:Nh}). Because the isolated-AGN subsample is small and selected differently from
the \citet{Ricci2017} samples, we do not interpret the black curve as a
formal match to any of their cumulative distributions over the full
\(N_{\rm H}\) range. Instead, we compare the high-column-density tail, quantified here by the
median column density and by the fractions of objects above
\(N_{\rm H}>10^{23}\,{\rm cm}^{-2}\) and
\(N_{\rm H}>10^{24}\,{\rm cm}^{-2}\). In this specific sense, the isolated AGNs occupy a heavily obscured
\(N_{\rm H}\) regime comparable to the high-\(N_{\rm H}\) end of the
late-stage merger AGNs distribution, although their cumulative curve remains
visibly distinct as a full distribution. This suggests that strong nuclear obscuration is not unique to galaxy
interactions and may also occur in isolated host galaxies, possibly due to
internal circumnuclear structures and/or inclination effects. Extending the sample of isolated AGNs and obtaining additional X-ray observations will be essential to quantify any environmental dependence of $N_{\rm H}$ while disentangling it from selection and geometry-related biases.

\subsection{Relation between the SMBH mass and X-ray luminosity }

Motivated by the regime-dependent behaviour of $\Gamma$--$l_X$ reported by \citet{2015MNRASYang}, we examined whether additional correlation features could be masked when analysing the full data set across the wide dynamic range of $l_X$.

\begin{figure}
    \centering
     \includegraphics[width=0.70\linewidth]{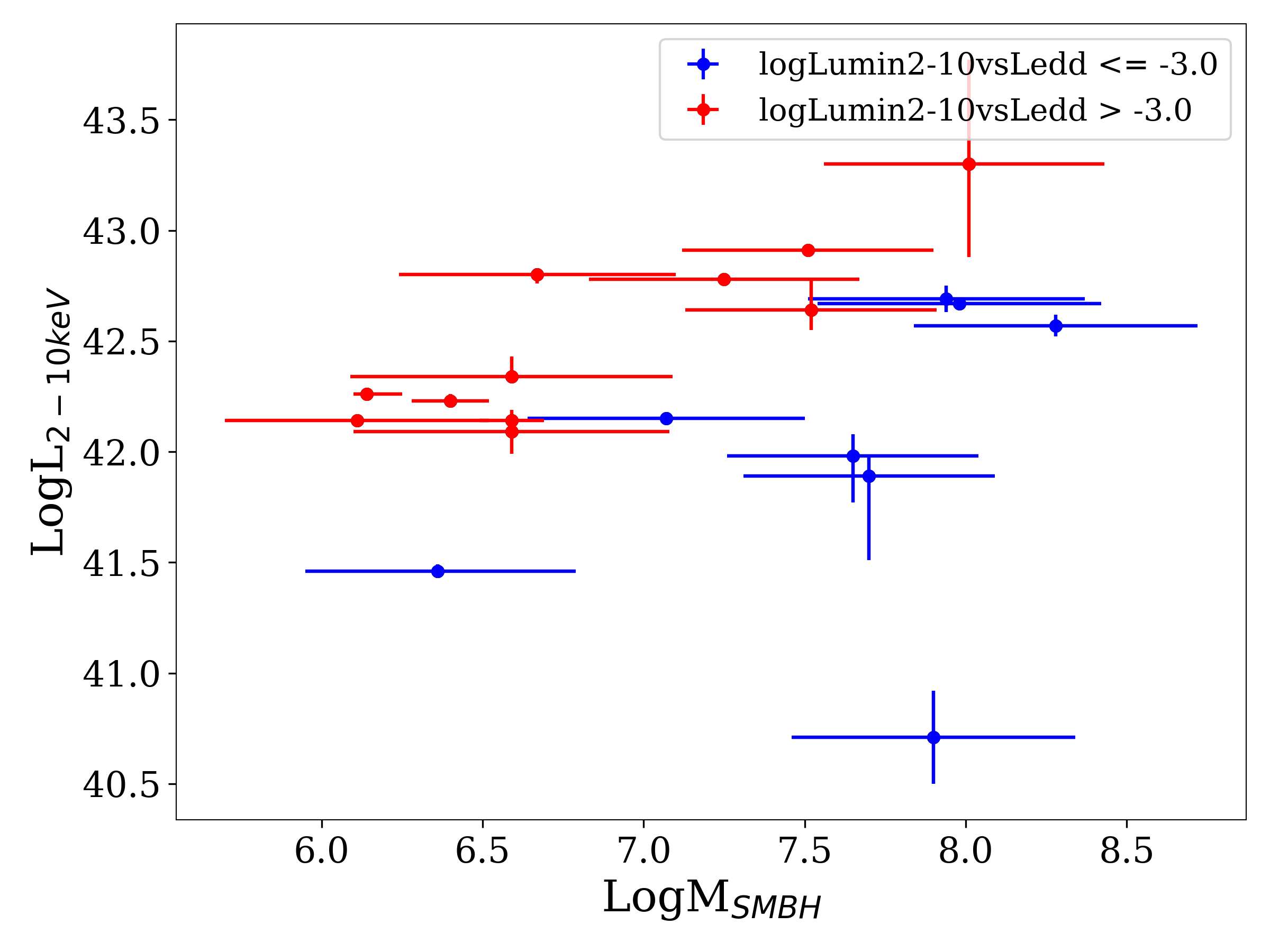}
    \includegraphics[width=0.72\linewidth]{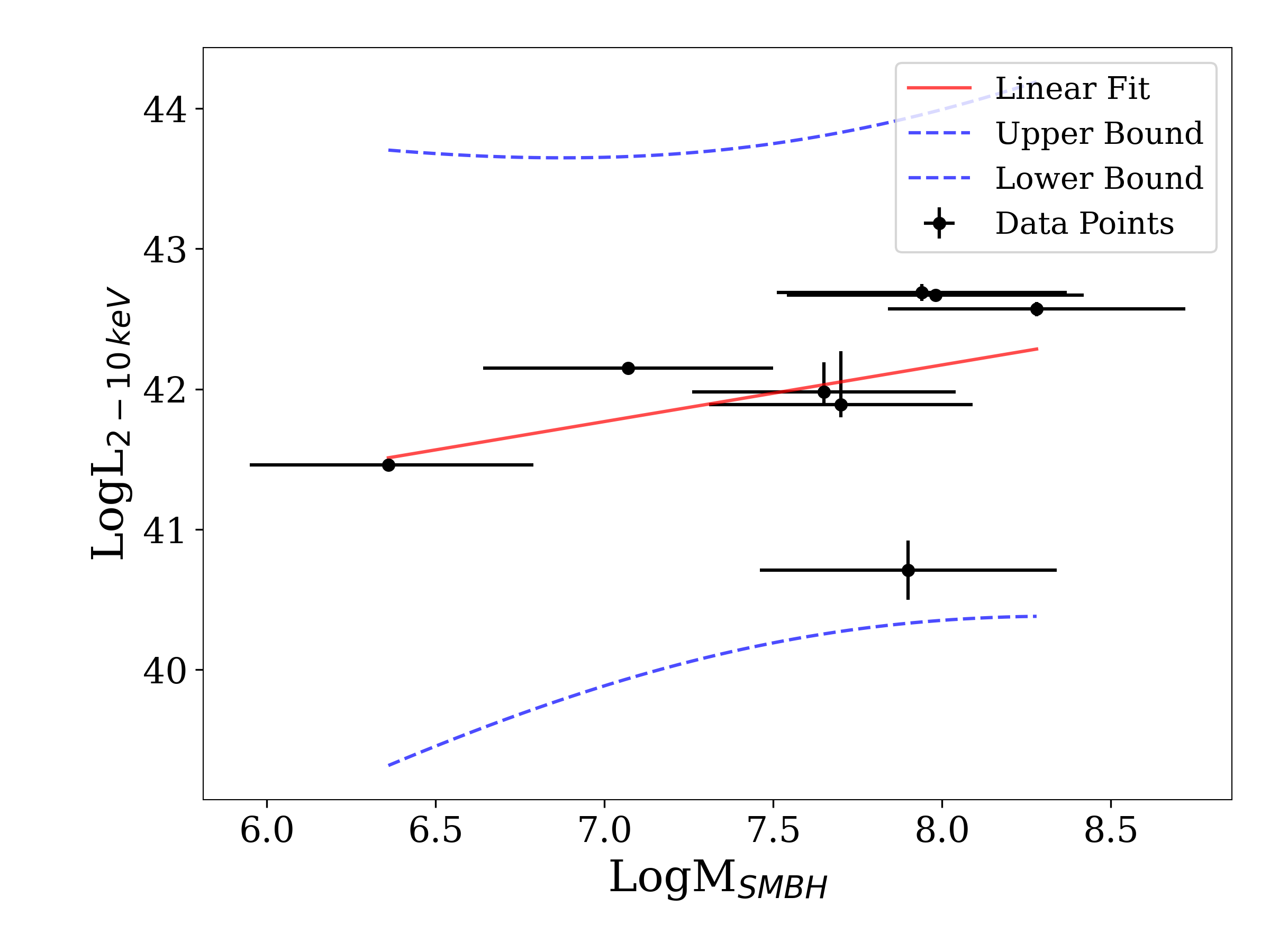}
     \includegraphics[width=0.72\linewidth]{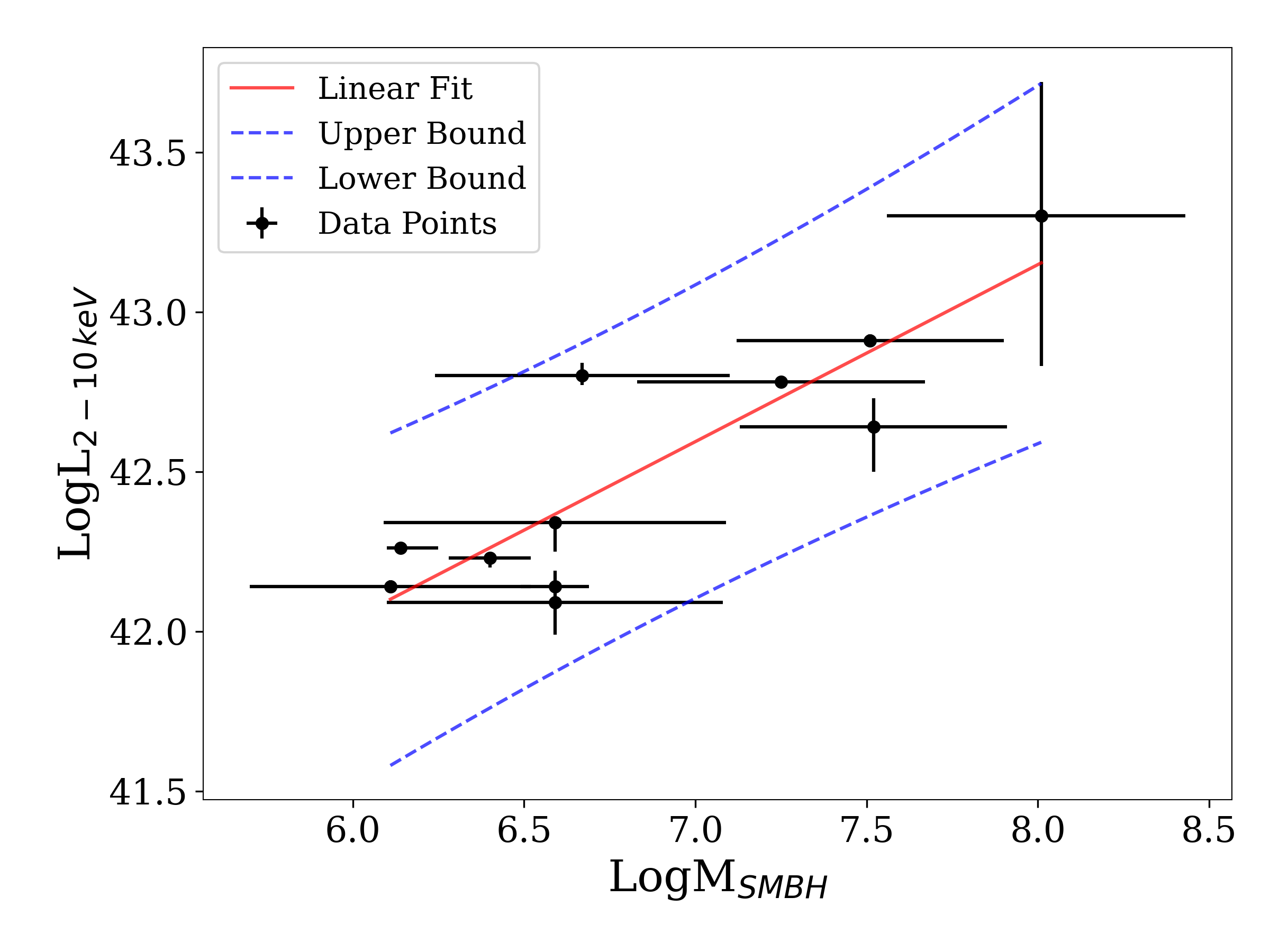}
  \caption{Relation between the X-ray intrinsic luminosity in the 2--10 keV band and SMBH masses for 2MIG isolated AGNs. Upper panel: the full sample; middle panel: the low-$l_X$ subsample ($l_X \leq 10^{-3}$); lower panel: the high-$l_X$ subsample ($l_X > 10^{-3}$).}
   \label{fig:L_MBH}
\end{figure}

After introducing the separation by accretion regime, we observe distinct behaviours for the low- and high-$l_X$ subsamples, where $l_X\equiv L_{2-10\,\mathrm{keV}}/L_{\mathrm{Edd}}$. For AGNs with $\log l_X \leq -3.0$, the uncertainty-aware Monte--Carlo Kendall coefficients remain moderate and statistically insignificant:
$\tau=-0.55$ ($p=0.066$) for the $\Gamma$--$\log\lambda_{\mathrm{Edd}}$ relation and
$\tau=-0.26$ ($p=0.47$) for $\Gamma$--$\log L_{2-10\,\mathrm{keV}}$,
suggesting no robust monotonic trends in the low-accretion regime.
For objects with $\log l_X>-3.0$, the $\Gamma$--$\log\lambda_{\mathrm{Edd}}$ pair remains statistically insignificant ($\tau=+0.35$, $p=0.19$), while the $\log L_{2-10\,\mathrm{keV}}$--$\log M_{\mathrm{SMBH}}$ relation becomes more coherent (see Fig.~\ref{fig:L_MBH}), motivating a dedicated regression analysis for this pair.

The performed linear regression analysis supports these findings. For the low-$l_X$ group ($\log l_X\le -3.0$), the best-fitting relation between luminosity and SMBH mass is
\[
\log L_{2-10\,\mathrm{keV}} = (0.4037 \pm 0.4208)\,\log M_{\mathrm{SMBH}} + (38.9432 \pm 3.2118),
\]
with a weak Pearson correlation coefficient $r=0.365$ and $p=0.375$.
In contrast, for the high-$l_X$ subset ($\log l_X>-3.0$), the slope becomes well constrained,
\[
\log L_{2-10\,\mathrm{keV}} = (0.5542 \pm 0.1052)\,\log M_{\mathrm{SMBH}} + (38.7137 \pm 0.7235),
\]
with $r=0.869$ and $p=5.14\times10^{-4}$, indicating a statistically significant positive correlation.

Finally, the Bayesian \texttt{linmix} regressions, which jointly account for measurement uncertainties in both axes and intrinsic scatter, yield consistent conclusions. While the full sample shows no statistically robust correlations (e.g., for $\log L_{2-10\,\mathrm{keV}}$ versus $\log M_{\mathrm{SMBH}}$, $\beta=0.162$ with 95\% CI $[-0.298,\,0.622]$ and $\sigma_{\rm int}\simeq0.64$),
the high-$l_X$ subsample exhibits a positive $L_{2-10\,\mathrm{keV}}$--$M_{\mathrm{SMBH}}$ relation with the posterior slope supported to be $>0$ (95\% CI $[0.168,\,0.941]$) and a substantially smaller intrinsic scatter ($\sigma_{\rm int}\simeq0.30$).
Taken together, the uncertainty-aware Monte--Carlo Kendall tests and the regression analyses indicate that the isolated-AGN population does not follow a universal global trend, whereas a tighter luminosity--mass sequence becomes visible for $\log l_X>-3$. However, this behaviour should be interpreted with caution (!). Since $l_X \equiv L_{2-10\,\mathrm{keV}}/L_{\mathrm{Edd}}$ and $L_{\mathrm{Edd}}\propto M_{\mathrm{SMBH}}$, the observed $L_{2-10\,\mathrm{keV}}$--$M_{\mathrm{SMBH}}$ trend in the high-$l_X$ subsample may be partly enhanced by the underlying scaling between luminosity, SMBH mass, Eddington ratio, and the adopted bolometric correction. Therefore, we do not interpret this result as evidence for a new independent physical correlation, but rather as a suggestive pattern that may reflect a narrower range of accretion states in this subsample. The limited sample size additionally cautions against overly strong generalisations.

Based on the results obtained in this study for isolated AGNs at $z < 0.05$, we conclude that the present X-ray data do not reveal any obvious spectral behaviour unique to isolated AGNs. This conclusion is supported by the presence of different accretion states and by the diversity of acceptable spectral models.
The spectral analysis (Section \ref{sec:Xray_generall}) shows that no particular basic/compound model has an advantage. The X-ray spectra of isolated AGNs are described by all types of models characteristic of AGN as such (Table \ref{tab:X-rayParam}). We have models applicable to CT Sy2, AGNs with a simple power-law spectrum, and even models in which it is probably necessary to account for relativistic effects. At the same time, estimates of intrinsic luminosity Log L$_{2-10\_keV}$ do not exceed the value $\sim 43$. Most of the isolated AGNs in our sample contain central black holes with masses $\geq 10^{7} M_{Sun}$. The SMBH masses of isolated AGNs fully correspond to the mass range of typical Seyfert galaxies and do not constitute, for example, a group of AGNs with some of the smallest SMBH masses. The presence of a luminosity threshold Log $L_{2-10\_keV} \leq 10^{43}$ is consistent with predominantly modest accretion power in the present sample. A tighter $L_{2-10\,\mathrm{keV}}$--$\log M_{\mathrm{SMBH}}$ sequence is seen only for the high-$l_X$ subsample, but this effect may be at least partly induced by the built-in dependence of $l_X$ on both luminosity and black hole mass, as well as by a comparatively narrow range of accretion states. Therefore, this trend cannot yet be considered a secure, independent signature of isolated AGNs and should be regarded as suggestive only until it is confirmed with a larger, more homogeneous sample.

\subsection{The Milky Way galaxies-analogues as the isolated AGNs}

Another intriguing aspect is the identification of potential Milky Way galaxy analogues (MWAs) based on morphological, structural and other properties. In this context for MWAs at the similar evolutionary stage, the morphological type should be the spiral galaxy with bar as well as such parameters as the isophotal and effective radii, luminosity, stellar mass, star formation rate, metallicity \citep{deVaucouleurs1978, Mutch2011, Licquia2015, Boardman2020a, Pilyugin2023, Tuntipong2024}, rotation curve \citep{McGaugh2016, Dmytrenko2023, Fedorov2025F} should be close to the same values of the Milky Way. Among other indicators for MWA search, we consider an isolation criteria, weak/absent nuclear activity, small SMBH mass of (1-9)$\times10^{6}M_{\odot}$, general multiwavelength properties in the sense of the similar shape of the spectral energy distribution \citep{2024KosNTVavilova, Kompaniiets2025a}. This advanced approach allows us to search for MWAs not only by certain selected properties, but also to refine the search for identifying Milky Way twins \citep{Kompaniiets2025}, taking in account that the Milky Way is an isolated galaxy with non-gravitationally significant satellites, which satisfies the environmental density conditions in the cosmic web \citep{Hirschmann2013}.  

 Regarding the SMBH-mass indicator, the nine galaxies in our study of X-ray properties of isolated AGNs at $z$ < 0.05 have $M_{\rm SMBH}\lesssim 9\times10^{6}\,M_{\odot}$, and four of them---UGC 10774, UGC~10120, NGC~6300, and CGCG~243--024---are classified as SAB, SBa/SBb (Table~\ref{tab:general}). We considered these objects as candidates for verifying other MWA selection criteria in isolated environments. An accretion state of three of them is Seyfert-like (Table \ref{tab:X-rayParam} and thus not representative of the current Milky Way nucleus. But UGC 10774 galaxy can be considered for future research as an MWA looking for its SAB morphology, LINER activity type, weak X-ray emission \citep{Nisbet2016}, and small SMBH mass (see Tables \ref{tab:general} and \ref{tab:SMBHmass}). 

\section{Summary}
\label{sec:conclusion}

We compiled all the available astroinformation from Swift, XMM-Newton, Chandra, NuSTAR, and INTEGRAL archives on the X-ray emission characteristics (Table \ref{tab:general}) and the SMBH masses of 2MIG isolated AGNs at $z$<0.05 (Table \ref{tab:X-rayParam} and Table \ref{tab:SMBHmass}). This enabled us to build spectral models and calculate SMBH masses and Eddington ratios for those isolated AGNs for which this had not previously been done (Table \ref{tab:X-rayParam}). All of the X-ray spectra were initially fitted with a simple power-law model. Furthermore, depending on the AGN type and the fit statistics, additional components of the X-ray spectrum were selected and included. 

We found that isolated AGNs do not tend to possess a preferential type of observed X-ray spectrum (Column ''Spectral model" in Table \ref{tab:X-rayParam}). There are spectra with a single simple component (e.g., ESO 215-014), with several simple components (e.g., NGC 5347, ESO 438-009), with an ionised reflection component (IGR J11366-6002, UGC 10120, NGC 5231), and with the multi-component (e.g., NGC 1050 and ESO 317-038) models. There are also seven objects (e.g., ESO 116-018, NGC 6300) that show the presence of a reprocessed emission component from an inhomogeneous dusty torus (i.e., a clumpy torus). The diversity of best-fit models reflects the wide variety of accretion regimes present in isolated AGNs. In most cases, the intrinsic 2–10 keV luminosities do not exceed log L$_{2-10 keV}^{int} \simeq$ 43. A particularly interesting case is ESO 499-041, for which Chandra data reveal deviations from the continuum above 5 keV that can be tentatively attributed to a relativistic iron line.

We estimated SMBH masses primarily using the M–$\sigma$ relation, supplemented by literature values where stellar velocity dispersions were unavailable. The resulting SMBH mass distribution of 32 isolated AGNs is concentrated mostly above \(10^6\,M_\odot\), with the majority of objects (18/32) lying in the \(10^7\)–\(10^8\,M_\odot\) range. We note, a larger homogeneous sample of isolated AGNs will be required to determine whether this represents a general property of the isolated-AGNs population. Bolometric luminosities and Eddington ratios were derived using hard X-ray bolometric corrections. 

The parameters obtained allowed us to identify specific dependencies consistent with a general understanding of physical processes in AGNs. In particular, we considered the mass-luminosity relationship, the photon index vs. the ratio of 2–10 keV luminosity to the Eddington luminosity. We find no evidence that isolation suppresses or enhances any particular accretion mode, as isolated AGNs exhibit a broad range of spectral properties and models. For the first time, we identified a tentative linear correlation between intrinsic 2–10 keV luminosity and SMBH mass, a trend not commonly observed in AGN samples from denser environments. Although this relation may be affected by small-number statistics, it suggests a potentially distinctive evolutionary behaviour of isolated systems. 

\section{DATA AVAILABILITY}
The data underlying this article will be shared on reasonable request
to the corresponding author.

\section*{Acknowledgements}
This research is part of the project (ID 848) conducted in the frame of the EURIZON programme (the EU grant agreement No.871072). The work by Vasylenko A.A. and Vavilova I.B. is supported by the National Research Foundation of Ukraine (Project No. 2023.03/0188). We thank Prof. Johan Knapen and Prof. John Beckman (IAC, Spain), Prof. Reynier Peletier (University of Groningen, The Netherlands) and Prof. Marek Kowalski (DESY, Germany) for fruitful seminars and discussions in March 2025, when the results of this research were finalised. Kompaniiets O.V. thanks Prof. Carlo Baccigalupi (SISSA, Italy) for discussions and recommendations as well as notes the Kyiv School of Economy support through the ''Talents for Ukraine" programme. The authors express our sincere gratitude to the reviewer for helpful detailed remarks, which allowed us to present the results in a more complete form.

This research has made use of data and/or software provided by the HEASARC, XMM-Newton, Swift, NuSTAR, NED, and HyperLEDA. The SAO/NASA ADS was very helpful in our research.

\bibliographystyle{mnras}
\bibliography{example} 



\appendix


\bsp	
\label{lastpage}
\end{document}